\documentclass[aps,twocolumn,preprintnumbers,showpacs,showkeys,nofootinbib,
superscriptaddress  
]{revtex4}
\usepackage{epsfig}
\usepackage{amssymb,amsmath,amsfonts,amsthm,graphicx,psfrag}
\graphicspath{{./Fig/}}                 % location for color  fig.
\newcommand{\beq}{\begin{equation}}
\newcommand{\eeq}{\end{equation}}
\newcommand{\bea}{\begin{eqnarray}}
\newcommand{\eea}{\end{eqnarray}}
\newcommand{\beas}{\begin{eqnarray*}}
\newcommand{\eeas}{\end{eqnarray*}}
\newcommand{\Fig}[1]{Fig.~\ref{#1}}
\newcommand{\Tab}[1]{Table~\ref{#1}}
\newcommand{\Sec}[1]{Section~\ref{#1}}
\newcommand{\Eq}[1]{Eq.~(\ref{#1})}
\newcommand{\re}{\operatorname{\mathfrak{Re}}}
\newcommand{\tr}{\operatorname{Tr}}

\newcommand{\bc}{{\it bc}}
\newcommand{\fc}{{\it fc}}
\newcommand{\LHS}{l.h.s.}
\newcommand{\RHS}{r.h.s.}

\newcommand{\aleq}{\mbox{}_{\textstyle \sim}^{\textstyle < }}
\newcommand{\ageq}{\mbox{}_{\textstyle \sim}^{\textstyle > }}

\begin{document}
\preprint{HU-EP-11/37, ITEP-LAT/2011-07}

\title{Landau gauge gluon and ghost propagators at finite temperature 
from quenched lattice QCD}
\author{R.~Aouane}
\affiliation{Humboldt-Universit\"at zu Berlin, Institut f\"ur Physik, 
12489 Berlin, Germany}
\author{V. Bornyakov}
\affiliation{Institute for High Energy Physics, 142281, Protvino, Russia \\
and Institute of Theoretical and Experimental Physics, 117259 Moscow, Russia}
\author{E.-M.~Ilgenfritz}
\affiliation{Humboldt-Universit\"at zu Berlin, Institut f\"ur Physik, 
12489 Berlin, Germany \\
and Joint Institute for Nuclear Research, VBLHEP, 141980 Dubna, Russia}
\author{V. Mitrjushkin}
\affiliation{Joint Institute for Nuclear Research, BLTF, 141980 Dubna, Russia \\
and Institute of Theoretical and Experimental Physics, 117259 Moscow, Russia}
\author{M.~M\"uller-Preussker}
\affiliation{Humboldt-Universit\"at zu Berlin, Institut f\"ur Physik, 
12489 Berlin, Germany}
\author{A.~Sternbeck}
\affiliation{Universit\"at Regensburg, Institut f\"ur Theoretische Physik, 
93040 Regensburg, Germany}

\date{December 31, 2011}

\begin{abstract}
%---------------
The behavior of the Landau gauge gluon and ghost propagators is studied in 
pure $SU(3)$ gauge theory at non-zero temperature on the lattice. 
We concentrate on the momentum range $[0.6, 2.0]~\mathrm{GeV}$. 
For the longitudinal as well as for the transverse component of the gluon 
propagator we extract the continuum limit. We demonstrate the smallness of 
finite-size and Gribov-copy effects at temperatures close to the 
deconfinement phase transition at $T=T_c$ and within the restricted 
range of momenta. Since the longitudinal component $D_L(q)$ turns out to be most 
sensitive with respect to the phase transition we propose some combinations 
of $D_L(q)$ signalling the transition much like ``order parameters''.    
\end{abstract}



\maketitle

\section{Introduction}
\label{sec:introduction}
%---------------------

It is commonly believed that hadronic matter at high temperature undergoes a 
phase transition into another phase, traditionally called ``quark-gluon plasma''. 
At present, strong efforts are made at RHIC, BNL and at the LHC, CERN to establish  
undeniable experimental signatures in the final states of heavy-ion collisions 
indicating that matter had undergone evolution close to or beyond this transition. 
The existence of such a transition has been concluded long time ago from
Hagedorn's thermodynamical model \cite{Hagedorn:1970gh}
and later has been one of the first crucial forecasts of lattice QCD (LQCD).
The latter uses a formulation of non-Abelian gauge theory which is amenable to
ab-initio numerical non-perturbative computations. This formulation also 
opens the way for analytical calculations at strong and weak
coupling. LQCD calculations can provide estimates for the transition 
temperature, the equation of state close to $T_c$ and above 
and other features and experimental observables.
For a recent review see \cite{DeTar:2011nm}.

In recent years, another powerful non-perturbative approach has been developed
based on Dyson-Schwinger equations (DSE) 
\cite{vonSmekal:1997vx,Hauck:1998fz, Roberts:2000aa,Maris:2003vk} and functional 
renormalization group equations (FRGE) \cite{Gies:2002af,Pawlowski:2003hq}. 
The main focus was first to find a field theoretical, model-independent description 
of quark and gluon confinement in terms of the infrared behavior of gauge-variant
Green's functions, in particular of the Landau or Coulomb gauge gluon and ghost 
propagators. This should allow to confirm or disprove confinement scenarios as 
proposed by Gribov and Zwanziger \cite{Gribov:1977wm,Zwanziger:2001kw,
Zwanziger:2003cf} and Kugo and Ojima \cite{Kugo:1979gm,Kugo:1995km}. Landau 
gauge gluon and ghost propagators have been intensively studied for zero 
temperature with DSE and FRGE (see, e.g., \cite{Fischer:2008uz} and citations 
therein). On the lattice these propagators have been computed by several groups 
(see \cite{Bornyakov:2009ug,Bogolubsky:2009dc} for our own recent computations 
and references to earlier work by other groups).

If these propagators encode confinement, they should also be
considerend in LQCD studies at non-vanishing temperature
(see \cite{Heller:1995qc,Heller:1997nqa,Cucchieri:2000cy,Cucchieri:2001tw} 
and for more recent work \cite{Cucchieri:2007ta,Maas:2009fg,Fischer:2010fx,
Cucchieri:2011di}). Complementary to this the temperature dependence has 
also been studied in the framework of DSE 
\cite{Gruter:2004bb,Maas:2005hs,Braun:2007bx,Fischer:2009wc}.

In the recent past we also have extended our lattice computations of the
gluon and ghost propagators at zero temperature to the case of non-zero
temperature. First results have been obtained without \cite{Bornyakov:2010nc}
and also with $N_f=2$ dynamical fermion flavors \cite{Bornyakov:2011jm}. 
In this paper we will focus on results for pure SU(3) gauge theory known
to have a first order finite temperature phase transition. 
We concentrate on the continuum limit within a restricted range of momenta. 
For this range finite-size or Gribov-copy effects turn out to be 
small. Moreover, a noticeable sensitivity of the longitudinal component 
of the gluon propagator with respect to the deconfining phase transition 
is observed. We show that certain ratios of this component 
may serve as useful indicators (``order parameters'') for this transition, 
thus complementing the information obtained from the ever popular 
Polyakov loop.

The paper is organized as follows. In \Sec{sec:setup} we describe the setup of 
our lattice Monte Carlo simulation. In \Sec{sec:propagators} we review the 
basic definitions of the gauge-variant propagators on the lattice, 
modified to finite temperature. In \Sec{sec:results} we present results 
for the gluon and ghost propagators for various temperatures. 
The signal of the phase transition is not as strong as one might have expected.
Nevertheless, as it is said above, the longitudinal gluon propagator allows us 
to define ratios which give a clear signal at the deconfinement phase transition. 
In Sections \ref{sec:finitevolume} and \ref{sec:gribov} we analyze the 
finite-volume and Gribov-copy effects, respectively. 
In \Sec{sec:scaling} we investigate then scaling properties for varying lattice 
spacing $a$, keeping the temperature and the volume fixed. This allows
us to extrapolate our data to the continuum limit.
Finally, in \Sec{sec:conclusions}, we shall draw our conclusions.

\section{Setup of the lattice simulations}
\label{sec:setup}
%-----------------------------------------

We have generated $SU(3)$ pure gauge field configurations on a four-dimensional 
lattice of size $N_{\sigma}^3 \times N_{\tau}$ with periodic boundary conditions 
employing standard Monte Carlo simulations using 
the Euclidean path integral weight $\sim\exp{(-S_W)}$, where $S_W$ denotes the 
Wilson one-plaquette action
\bea
S_W&=&\beta \sum_{x; \mu >\nu}
\left[ 1 -\frac{1}{3} \re\tr \Bigl(U_{x\mu}U_{x+\hat{\mu};\nu}
U_{x+\hat{\nu};\mu}^{\dagger}U_{x\nu}^{\dagger} \Bigr)\right],\nonumber \\
&&\beta = 6/g_0^2 \, . \nonumber
\label{action}
\eea    
$g_0$ is the bare coupling constant and $U_{x\mu} \in SU(3)$ denotes the link variables. 
The imaginary-time extent corresponds to the inverse temperature $T^{-1}=N_{\tau} a$, 
where $a(\beta)$ is the lattice spacing. 
For generating the gauge field ensemble we have used the standard hybrid
over-relaxation algorithm, with a step of 4 microcanonical over-relaxation sweeps 
followed by one heatbath step \cite{Fabricius:1984wp,Kennedy:1985nu}. In
both steps a decomposition of SU(3) link 
variables into SU(2) matrices, as proposed in \cite{Cabibbo:1982zn},
was applied. $O(2000)$ combined thermalization sweeps were allowed between
the individual measurements of the propagators. 

%------------------------------------------------------------------------------
\begin{table}
\centering
\begin{tabular}{|c|c|c|c|c|c|c|c|}
\hline
$T/T_{c}$ & $N_{\tau}$ & $N_{\sigma}$ & $\beta$ & $a(\mathrm{GeV}^{-1})$ 
                       & ~$a(\mathrm{fm})$~& $n_{conf}$  & $n_{copy}$\\
\hline
0.65 & 18 & 48 & 6.337 & 0.28 & 0.055 & 150 & 1  \\
0.74 & 16 & 48 & 6.337 & 0.28 & 0.055 & 200 & 1  \\
0.86 & 14 & 48 & 6.337 & 0.28 & 0.055 & 200 & 1  \\
0.99 & 12 & 48 & 6.337 & 0.28 & 0.055 & 200 & 1  \\
1.20 & 10 & 48 & 6.337 & 0.28 & 0.055 & 200 & 1  \\
1.48 &  8 & 48 & 6.337 & 0.28 & 0.055 & 200 & 1  \\
1.98 &  6 & 48 & 6.337 & 0.28 & 0.055 & 200 & 1  \\
2.97 &  4 & 48 & 6.337 & 0.28 & 0.055 & 210 & 1  \\
\hline\hline                                         
0.86 &  8 & 28 & 5.972 & 0.49 & 0.097 & 200 & 27 \\
0.86 & 12 & 41 & 6.230 & 0.33 & 0.064 & 200 & 1  \\
0.86 & 16 & 55 & 6.440 & 0.24 & 0.048 & 200 & 1  \\
\hline
1.20 &  6 & 28 & 5.994 & 0.47 & 0.094 & 200 & 27 \\
1.20 &  8 & 38 & 6.180 & 0.35 & 0.069 & 200 & 1  \\
1.20 & 12 & 58 & 6.490 & 0.23 & 0.045 & 200 & 1  \\
\hline\hline                                            
0.86 & 14 & 56 & 6.337 & 0.28 & 0.055 & 200 & 1  \\
0.86 & 14 & 64 & 6.337 & 0.28 & 0.055 & 200 & 1  \\
\hline
1.20 & 10 & 56 & 6.337 & 0.28 & 0.055 & 200 & 1  \\
1.20 & 10 & 64 & 6.337 & 0.28 & 0.055 & 200 & 1  \\
\hline
\end{tabular}
\caption{Temperature values, lattice size parameters, values of the inverse
bare coupling $\beta$, the lattice spacing $a$ in units of $\mathrm{GeV}^{-1}$ 
and $\mathrm{fm}$, the number $n_{conf}$ of independent lattice field 
configurations and the number $n_{copy}$ of gauge copies used throughout this 
study.}
\label{tab:numbers}
\end{table}
%------------------------------------------------------------------------------

In order to determine the temperature dependence of the gluon and
ghost propagators, in a first step we kept the lattice spacing fixed 
(and, as we shall see, sufficiently small) while varying $N_{\tau}$. 
As a reference value we have chosen $\beta=6.337$ providing 
$a \simeq 0.055~\mathrm{fm}$ (in accordance with \cite{Necco:2001xg}). 
This $\beta$-value corresponds, for $N_{\tau}=12$, to a temperature 
very close to the temperature $T_c$ characteristic for the deconfinement 
phase transition in a lattice with a linear spatial extent 
$N_{\sigma} a(\beta=6.337) = 48~a~\simeq 2.64~\mathrm{fm}$. 
According to Ref.~\cite{Boyd:1996bx} it has been fixed by interpolating 
with the help of the fit formula   
\beq \nonumber
\beta_{c}(N_{\tau},N_{\sigma})=\beta_{c}(N_{\tau},\infty)
                     - h \left(\dfrac{N_{\tau}}{N_{\sigma}}\right)^3\,,
\eeq 
where $\beta_{c}(N_{\tau},\infty)$ corresponds to the thermodynamic 
limit and $h$ denotes a fitted coefficient ($h \lesssim 0.1$). 
$N_{\sigma} = 48$ guarantees a reasonable aspect ratio over the whole 
temperature range $T/T_c \equiv 12/N_{\tau} \in [12/18, 12/4]$ and 
permits to reach three-momenta below $1~\mathrm{GeV}$.

As a second step, we decided to study systematic effects as there are 
finite-volume effects (cf. \Sec{sec:finitevolume}), Gribov copy effects 
(cf. \Sec{sec:gribov}) and the scaling properties (cf. \Sec{sec:scaling}) 
in order to extrapolate to the continuum limit $a \to 0$ for a couple of 
momentum values. For the two latter studies we varied $a(\beta)$ while 
having kept constant the physical spatial volume $(2.7~\mathrm{fm})^3$ 
as well as two representative temperature values ($T \simeq 0.86 \,T_c$  
and $T \simeq 1.20 \,T_c$, respectively).
  
A compilation of the lattice sizes $(N_{\tau} \times N_{\sigma}^3)$ and 
$\beta$-values together with the number of independent lattice configurations 
generated for this study can be found in \Tab{tab:numbers}.

\section{Gluon and ghost propagators}
\label{sec:propagators}
%------------------------------------

For determining the gluon and ghost propagators we have to fix the gauge.
Under local gauge transformations $\{g_x\}$ the link variables transform as 
\beq
U_{x\mu} \stackrel{g}{\mapsto} U_{x\mu}^{g}
= g_x^{\dagger} U_{x\mu} g_{x+\hat{\mu}} \,,
\qquad g_x \in SU(3) \,.
\label{gaugetraf}
\eeq
In order to satisfy the Landau gauge transversality condition
\beq 
 \nabla_{\mu}A_{\mu}=0
\label{eq:gaugecondition}
\eeq
with the lattice gauge potentials
\beq
 A_{\mu}(x+\hat{\mu}/2)=
               \frac{1}{2iag_{0}}(U_{x\mu}-U_{x\mu}^{\dagger})\mid_{traceless}
\label{eq:potential}
\eeq
it is sufficient to maximize the gauge functional
\beq
F_{U}[g]=\dfrac{1}{3} \sum_{x,\mu} \re \tr g_{x} U_{x\mu} g_{x+\hat{\mu}}^{\dagger}
\label{eq:functional}
\eeq
with respect to $g_{x}$. What concerns the Gribov non-uniqueness problem for 
solutions of the gauge condition \Eq{eq:gaugecondition} we adopt the 
strategy of finding gauge copies being as close as possible to the global 
maximum of $F_{U}[g]$ \cite{Parrinello:1990pm,Zwanziger:1990tn} as already 
practized in \cite{Bakeev:2003rr,Sternbeck:2005tk,Bogolubsky:2005wf,
Bogolubsky:2007bw,Bornyakov:2009ug,Bogolubsky:2009dc}.
This prescription has been shown to provide correct results for Landau gauge
photon and fermion propagators within compact $U(1)$ lattice gauge theory 
\cite{Nakamura:1991ww,Bornyakov:1993yy,Bogolubsky:1999cb,Bogolubsky:1999ud}. 
Very efficient for this aim is the simulated annealing (SA) algorithm combined 
with subsequent overrelaxation (OR) iterations 
\cite{Bali:1994jg,Bali:1996dm,Bogolubsky:2007pq,Bornyakov:2009ug,Bogolubsky:2009dc}. 
The SA algorithm generates gauge transformations $\{g_{x}\}$ randomly with a 
statistical weight $\sim \exp(F_{U}[g]/T_{sa})$. The 
``temperature'' $T_{sa}$ is a technical parameter 
which is monotonously lowered in the course of $3500$ SA simulation sweeps 
(actually, these are heatbath updates). Also, for better performance, a few 
microcanonical steps are applied after each heatbath step. In fact, we start 
with $T_{sa}=0.45$ and decrease this parameter down to 
$T_{sa}=0.01$ in equal steps after each combined sweep. Finally, in order to 
satisfy the gauge condition \Eq{eq:gaugecondition} with a local accuracy of
\beq
\max_{x}\re\tr[\nabla_{\mu}A_{x\mu}\nabla_{\nu}A_{x\nu}^{\dagger}]<\varepsilon\,,
\quad \varepsilon=10^{-13}
\eeq
we employ the standard OR procedure. Except for the study of the influence of 
Gribov copies (cf.\ \Sec{sec:gribov}), we carry out only one such attempt per 
configuration to fix the gauge. As in our previous studies, we call the 
corresponding (first trial) gauge copy ``first copy'' (\fc).

The gluon propagator is defined in momentum space as
\beq
 D^{ab}_{\mu\nu}(q) 
 =\left\langle \widetilde{A}^a_{\mu}(k)\widetilde{A}^b_{\nu}(-k) \right\rangle,
\label{eq:gluonzero}
\eeq
where $\langle\cdots\rangle$ represents the average over configurations, and 
$\widetilde{A}^a_{\mu}(k)$ denotes the Fourier transform of the 
gauge-fixed gluon field (\ref{eq:potential}) depending on the integer-valued 
lattice momentum $k_{\mu}$ ($\mu=1,\ldots,4$). The latter is related to the 
physical momentum (for the Wilson plaquette action) as 
\beq
q_{\mu}(k_{\mu}) = \frac{2}{a} \sin\left(\frac{\pi
 k_{\mu}}{N_{\mu}}\right),~\quad 
 k_{\mu} \in \left(-N_{\mu}/2, N_{\mu}/2\right],
\eeq
where $(N_i, i=1,2,3; N_4) \equiv (N_{\sigma}; N_{\tau})$ characterizes 
the lattice size.

For non-zero temperature it is convenient to split the propagator 
into two components, the transverse $D_T$ (``chromomagnetic'') (transverse to 
the heatbath rest frame) and the longitudinal $D_L$ one (``chromoelectric''), 
respectively,
\beq
D^{ab}_{\mu\nu}(q)=\delta^{ab} 
            (P^{T}_{\mu\nu} D_{T}(q_{4}^{2},\vec{q}^{\,2})+
                   P^{L}_{\mu\nu} D_{L}(q_{4}^{2},\vec{q}^{\,2})),
\eeq
where $q_4$ plays the r\^ole of the Matsubara frequency, which will be put to 
zero lateron.
For the Landau gauge, the tensor structures $P^{T,L}_{\mu\nu}$ represent 
projectors transverse and longitudinal relative to the $(\mu=4)$-direction
\bea
P^{T}_{\mu\nu}&=&(1-\delta_{\mu4})(1-\delta_{\nu4}) 
       \left(\delta_{\mu\nu}- \frac{q_{\mu}q_{\nu}}{\vec{q}^{\;2}}\right), \\
P^{L}_{\mu\nu}&=&\left(\delta_{\mu\nu}-\frac{q_{\mu}q_{\nu}}{\vec{q}^{\;2}}\right)
                -P^{T}_{\mu\nu}\,.
\eea
For the propagator functions $D_{T,L}$ we find
\beq
 D_T=\frac{1}{2 N_g} 
        \left\langle\sum_{i=1}^3  
         \widetilde{A}^a_i(k)\widetilde{A}^a_i(-k)
        -\frac{q_4^2}{\vec{q}^{\;2}} 
        \widetilde{A}^a_4(k)\widetilde{A}^a_4(-k)\right\rangle
\eeq
and
\beq
 D_L= \frac{1}{N_g}\left(1 + \frac{q_4^2}{\vec{q}^{\;2}}\right) 
        \left\langle \widetilde{A}^a_4(k) \widetilde{A}^a_4(-k) \right\rangle,
\eeq
where the number of generators $N_g=N_{\rm color}^2-1$ for $N_{\rm color}=3$.
The zero-momentum propagator values can be defined as
\bea
\label{eq:zeromomprop}
D_T(0) &=& \frac{1}{3 N_g}
     \sum_{i=1}^3 \left\langle \widetilde{A}^a_i(0) \widetilde{A}^a_i(0) \right\rangle, 
\\
D_L(0) &=& \frac{1}{N_g}
       \left\langle \widetilde{A}^a_4(0) \widetilde{A}^a_4(0) \right\rangle. 
\eea

Notice that -- at least for large enough $\beta$ -- the Landau gauge gluon 
propagator is expected to depend on the ${\bf Z}(3)$-sectors into which the 
Polyakov loop spatial averages can fall \cite{Damm:1998pd}. Therefore, before 
carrying out the SA gauge fixing procedure we always apply a ${\bf Z}(3)$-flip 
as described in \Sec{sec:gribov} but with respect to the 4-th direction. It ensures 
the phases of the corresponding Polyakov loop averages to fall into the 
interval $(-\pi/3, \pi/3]$.    

The Landau gauge ghost propagator $G(q)$ and its dressing function $J(q)$ are
defined as follows.
\bea
\label{eq:ghost} \nonumber
G^{ab}(q)&=&a^{2}\sum_{x,y}\langle e^{-2\pi i(k/N)\cdot(x-y)} [M^{-1}]^{ab}_{xy}\rangle, 
    \\
         &=&\delta^{ab}~G(q) \equiv \delta^{ab}~J(q)/q^2,
\eea

\noindent where $q^2 \ne 0$ and
$(k/N)\cdot(x-y) \equiv \sum_{\mu} k_{\mu} (x-y)_{\mu} / N_{\mu}$.
$M$ denotes the lattice Faddeev-Popov operator corresponding to the gauge field 
definition (\ref{eq:potential}) and the related gauge functional
(\ref{eq:functional}), i.e.,
\beq
M^{ab}_{xy}=\sum_{\mu}[A^{ab}_{x,y}\delta_{x,y}-
                       B^{ab}_{x,y}\delta_{x+\hat{\mu},y}-
                       C^{ab}_{x,\mu}\delta_{x-\hat{\mu},y}]
\eeq
with
\begin{align*}
 A^{ab}_{x,y}&=\quad \re \tr [\{T^{a},T^{b}\}(U_{x,\mu}+U_{x-\hat{\mu},\mu})], \\
 B^{ab}_{x,y}&=2\cdot \re \tr [T^{b} T^{a} U_{x,\mu}],\\
 C^{ab}_{x,y}&=2\cdot \re \tr [T^{a} T^{b} U_{x-\hat{\mu},\mu}],
\end{align*}
where $T^{a}$ ($a= 1, \ldots, N_g$) are the Hermitian generators of the 
\textit{su(3)} Lie algebra satisfying $\tr [T^{a} T^{b}]=\delta^{ab}/2$.
In order to invert $M$ we use the conjugate gradient (CG) algorithm with 
plane-wave sources $\vec{\psi}_{c}$ with color and position components 
$\psi^{a}_{c}(x)= \delta^{a}_{c}\exp\left(2\pi\,i (k/N)\cdot x\right)$. 
Actually, we apply a pre-conditioned CG algorithm to solve the equations 
$M^{ab}_{xy}\phi^{b}(y)=\psi^{a}_{c}(x)$, where as pre-conditioning matrix 
we use the inverse Laplacian $\Delta^{-1}$ with a color-diagonal substructure 
\cite{Sternbeck:2006cg,Sternbeck:2006rd}.

In order to study hypercubic lattice artifacts, we
have analyzed the influence of the choice of momenta on the behavior of the 
gluon propagator. 
When comparing on-axis with diagonal momenta (for $\beta=6.337$ and
in the lower momentum range) we found only small but nonetheless
systematic deviations due to the hypercubic lattice geometry. To
maximally reduce them the so-called cylinder cut \cite{Leinweber:1998uu}
\beq \label{eq:cylcut}
 \sum_\mu k^2_\mu - \frac{1}{4}(\sum_\mu k_\mu )^2 \leq c,
\eeq
with $k_4=0$ and $c=3$ has been applied to all our data.

\section{Results: Gluon and ghost propagators versus temperature}
\label{sec:results}
%---------------------------------------------------------------

\begin{figure*}[tb]
 \mbox{\hspace*{-0.5cm}
 \includegraphics[angle=0,scale=0.7]{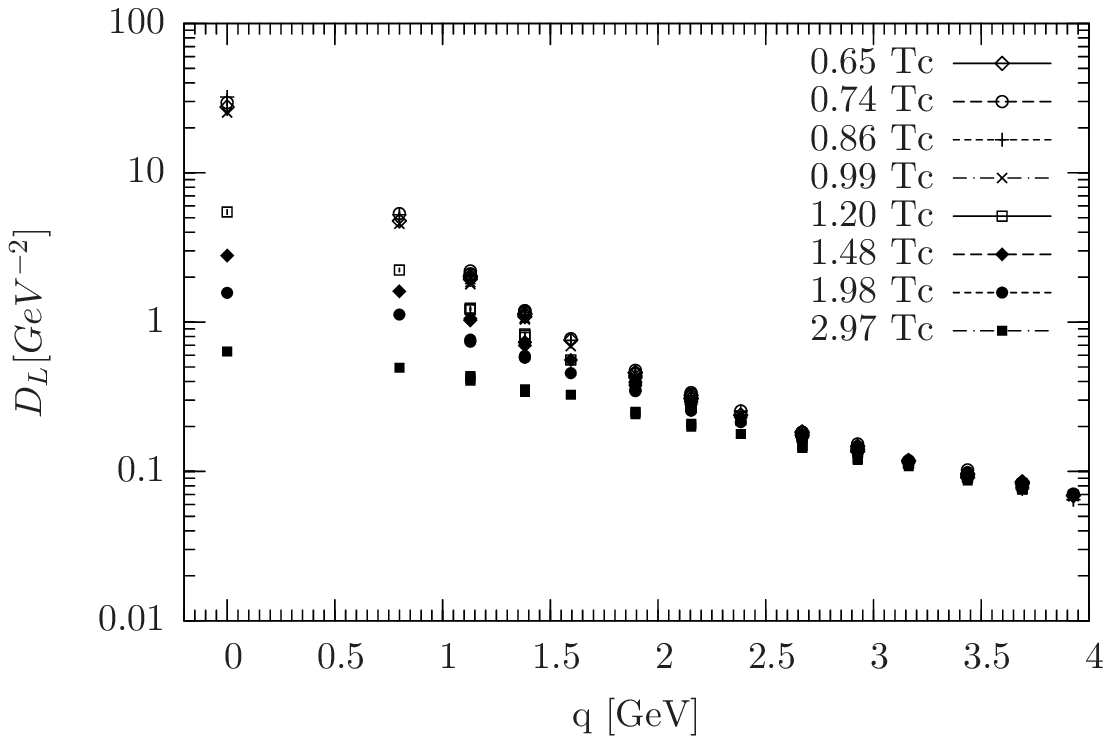} 
 \hspace*{-1.5cm}
 \includegraphics[angle=0,scale=0.7]{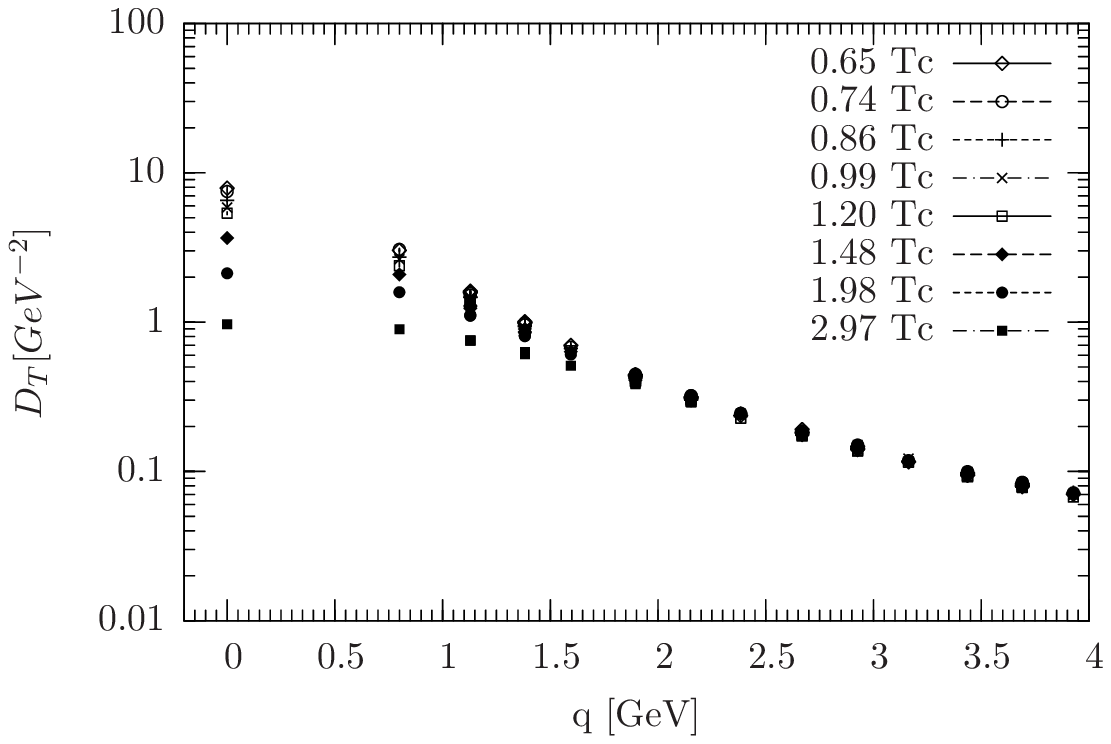} 
 }
\caption{Temperature dependence of the longitudinal (\LHS) and the 
transverse (\RHS) gluon propagator for $\beta=6.337$ and a spatial lattice 
size $N_{\sigma}=48$.}
\label{fig:templongtrans}
\end{figure*}

In \Fig{fig:templongtrans} we display the multiplicatively renormalized 
propagators $D_L(q)$ and $D_T(q)$ as functions of the three-momentum 
($q\equiv |\vec{q}|,~q_4=0$) for $\beta=6.337$, obtained with $N_{\sigma}=48$ 
and different $N_{\tau}$, i.e. for temperature values varying from $T=0.65\,T_c$ 
up to $T \simeq 3\,T_c$. For details we refer to the upper section of \Tab{tab:numbers}.
The renormalization condition is chosen such that $D_{L,T}$ take their tree level 
values at the subtraction point $q=\mu$. We choose $\mu=5~\mathrm{GeV}$ in 
order to be close to the perturbative range and still reasonably away from our 
lattice cutoff ($q_{max}= 2 \sqrt{3}/a \simeq 12.4~\mathrm{GeV}$).

\begin{figure*}[tb]
 \mbox{\hspace*{-0.2cm}
 \includegraphics[angle=0,scale=0.7]{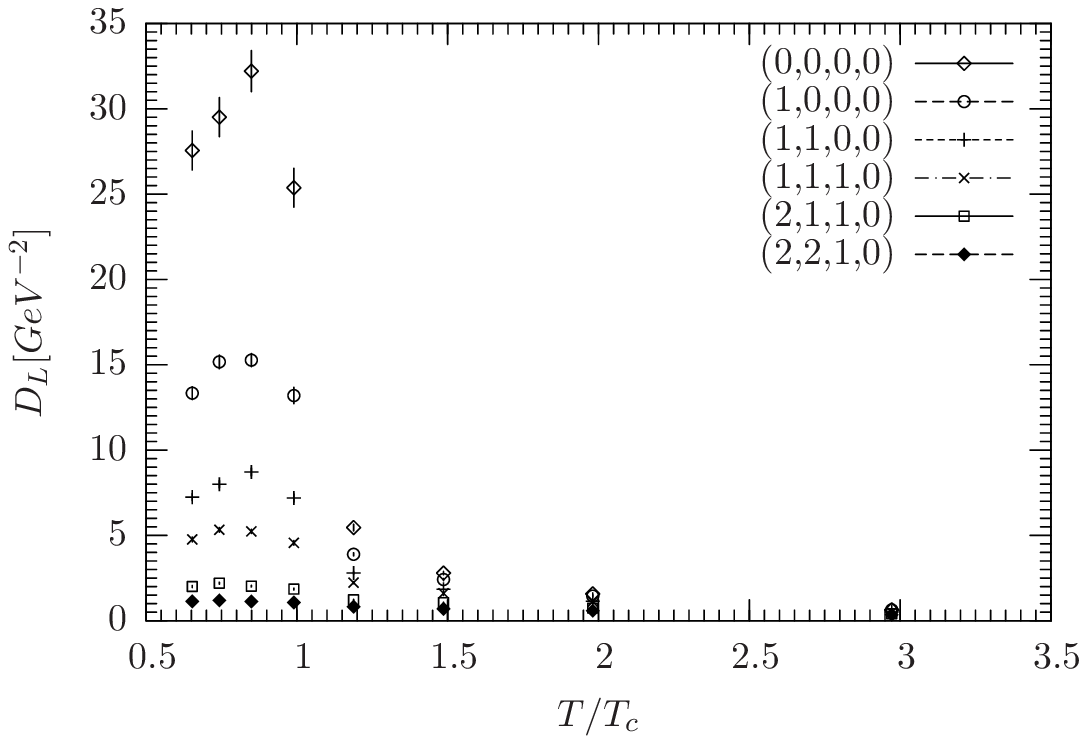} 
 \hspace*{-1.0cm}
 \includegraphics[angle=0,scale=0.7]{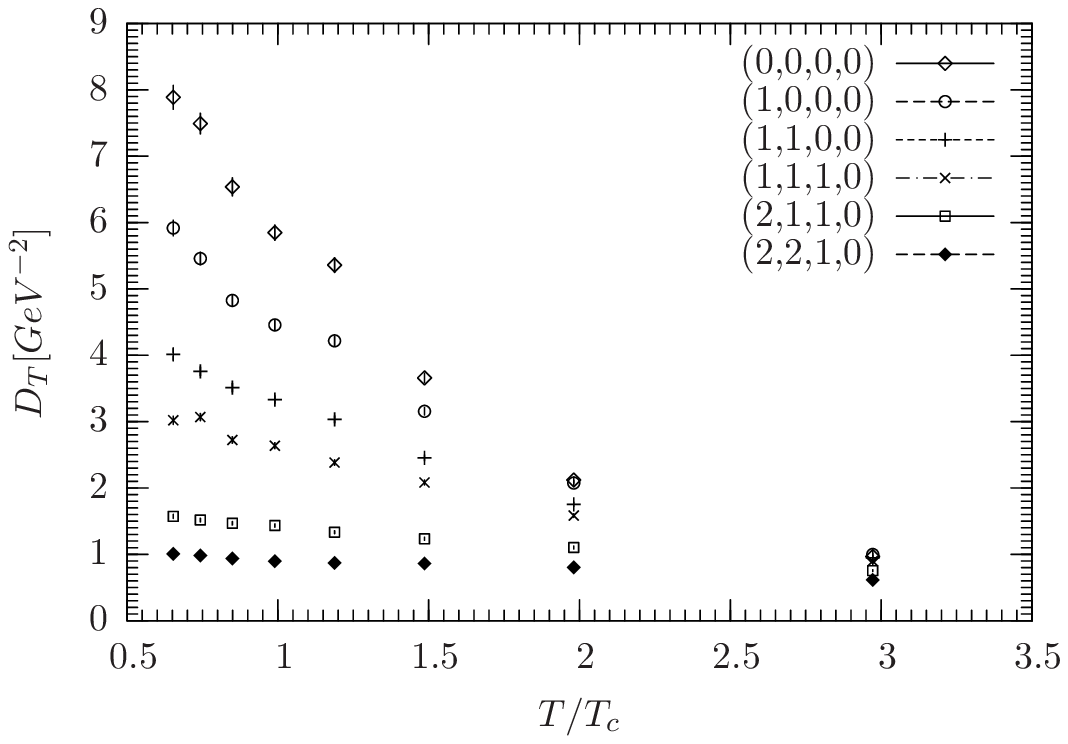} 
 }
\caption{The longitudinal propagator, $D_L$, (\LHS) and the transverse one, 
$D_T$, (\RHS) vs.\ temperature for a few low momenta, the latter represented 
as $(k_1,k_2,k_3,k_4)$. $\beta=6.337$ and $N_{\sigma}=48$. 
}
\label{fig:templongtranslow}
\end{figure*}

One can see from \Fig{fig:templongtrans} that the temperature dependence of 
both $D_L$ and $D_T$  becomes weaker with increasing momentum. 
This weakening proceeds faster for $D_T$ than for $D_L$. 
The ultraviolet regions of $D_T$ and $D_L$ turn out to be ``phase-insensitive''. 
This observation was also reported in~\cite{Bornyakov:2011jm}.
More precisely, while the temperature changes from its minimal value to 
our maximal one, the change of $D_T$ is less than 5\% for $q > 2.2 $ GeV, 
while for $D_L$ this is guaranteed for $q > 2.7 $ GeV.
For $~T~\aleq~T_c$ $~D_L~$ shows a comparatively weak temperature dependence 
also at small momenta. This changes drastically as soon as $T~\ageq~T_c$. 
In contrast to that $D_T(q)$ changes monotonously with $T$ in the 
infrared region.
This can be seen in more detail from \Fig{fig:templongtranslow}. There
we show the temperature dependence of $D_L(q)$ (left panel) as well as of 
$D_T(q)$ (right panel) for six selected momenta in the range up to 
$1.6~\mathrm{GeV}$. 

%------------------------------------------------------------------------------
%------------------------------------------------------------------------------
%------------------------------------------------------------------------------
%------------------------------------------------------------------------------
One can see that $D_L$ at fixed momentum shows strong variations in the 
neighbourhood of $T_{c}$. It is rising with $T$ 
below $T_{c}$ and sharply drops around $T_c$. This behavior looks most 
pronounced for zero momentum and gets progressively weaker at higher 
momenta. For the lowest momenta we observe maxima at $T=0.86\,T_c$. 
It remains open, whether the maxima are shifted away from the transition 
temperature with increasing volume.% 
\footnote{For $SU(2)$ gauge theory the maximum of $D_L(0)$ was 
recently reported~\cite{Cucchieri:2011di} to move away from
the transition with decreasing lattice spacing.}

In any case, our data confirms that the infrared part of $D_L(p)$ is 
strongly sensitive to the temperature phase transition 
\cite{Fischer:2010fx,Bornyakov:2011jm}. 
It may serve to construct some kind of order parameters characterizing the 
onset of the phase transition, as we will propose below. In contrast to that, 
$D_T$ is ever decreasing and varying smoothly across $T_{c}$, showing no 
visible response to the phase transition at all momenta. 

We fit the momentum dependence in the range 
$[0.6 : 8.0]~\mathrm{GeV}$ with a Gribov-Stingl interpolation 
formula \cite{Gribov:1977wm,Stingl:1994nk} 
used in Refs. \cite{Cucchieri:2003di,Cucchieri:2011di} 
and derived lateron in the so-called ``Refined Gribov-Zwanziger'' 
approach \cite{Dudal:2008sp,Dudal:2011gd}
\beq \label{eq:stinglfit}
D(q)=\frac{c~(1+d\,q^{2n})}{(q^2+r^2)^2+b^2}\,.
\eeq
Expected logarithmic corrections needed for the ultraviolet limit have been 
neglected here (for a thorough discussion see \cite{Leinweber:1998uu}).
We put throughout $n=1$. In a first attempt we have left $b$ varying. 
We obtained values compatible with $b\,=\,0$ except for $D_T(q)$ at the highest 
three temperature values inspected. Therefore, in all other cases we have
repeated the fits with fixed $b=0$ and obtained $\chi^2_{df}$-values 
reasonably below $\,2.0$. The fit parameters can be found in \Tab{fit:DallT}.% 
\footnote{Note that for $b=0~$ \Eq{eq:stinglfit} is equivalent to
the interpolation formula
$~D(q)=\frac{\gamma}{(q^2+\delta^2)}+\frac{\beta}{(q^2+\delta^2)^2}\,.$}
Since we expect to see a plateau and even a bend over for $D_T$ 
at momenta below our minimal ones the parameter $b$ might become nonzero
also at lower temperatures. This would then correspond to a complex 
effective mass parameter.  

%------------------------------------------------------------------------------
\begin{table*}
\mbox{ 
\setlength{\tabcolsep}{1.0pt}
\begin{tabular}{|c|c|c|c|c|c|c|} 
\hline
\multicolumn{2}{|c|}{Parameters} & \multicolumn{5}{|c|}{$D_L$ fits} \\
\hline
 $~T/T_{c}~$ & $~N_{\tau}~$ & $~r^2(\mathrm{GeV}^{2})~$ & $~b(\mathrm{GeV}^{2})~$ 
      & $d(\mathrm{GeV}^{-2})$ & $c(\mathrm{GeV}^{2})$ & $\chi_{df}^{2}$ \\
\hline
0.65 & 18 & 0.372(29) & {\bf 0.0} & 0.192(8) & 4.29(17) & 1.49 \\
0.74 & 16 & 0.296(22) & {\bf 0.0} & 0.206(7) & 4.11(13) & 1.40 \\
0.86 & 14 & 0.257(22) & {\bf 0.0} & 0.221(8) & 3.70(13) & 1.57 \\
0.99 & 12 & 0.359(30) & {\bf 0.0} & 0.209(10) & 3.89(16) & 1.83 \\
1.20 & 10 & 1.029(41) & {\bf 0.0} & 0.155(6) & 5.43(21) & 1.27 \\
1.48 &  8 & 1.547(47) & {\bf 0.0} & 0.118(4) & 7.12(24) & 1.06 \\
1.98 &  6 & 2.455(75) & {\bf 0.0} & 0.086(4) & 9.55(37) & 1.35 \\
2.97 &  4 & 5.327(159) & {\bf 0.0} & 0.045(2) & 17.15(73) & 0.51 \\
\hline 
\end{tabular}
}
\mbox{
\setlength{\tabcolsep}{1.0pt}
\begin{tabular}{|c|c|c|c|c|}  
\hline
\multicolumn{5}{|c|}{$D_T$ fits} \\
\hline
 $~r^2(\mathrm{GeV}^{2})~$ & $~b(\mathrm{GeV}^{2})~$ & $d(\mathrm{GeV}^{-2})$ 
                           & $c(\mathrm{GeV}^{2})$ & $\chi_{df}^{2}$ \\
\hline
 0.751(24) &  {\bf 0.0} & 0.153(4) & 5.40(14) & 1.17 \\
 0.756(20) &  {\bf 0.0} & 0.161(3) & 5.31(11) & 0.99 \\
 0.847(22) &  {\bf 0.0} & 0.152(4) & 5.50(12) & 1.09 \\
 0.869(26) &  {\bf 0.0} & 0.157(4) & 5.45(14) & 1.44 \\
 0.951(25) &  {\bf 0.0} & 0.147(4) & 5.56(13) & 1.17 \\
 0.886(138) & 0.810(167) & 0.146(11) & 5.70(42) & 1.46 \\
 0.856(109) & 1.398(62) & 0.133(8) & 6.15(34) & 0.93 \\
 0.927(126) & 2.559(33) & 0.100(6) & 7.58(41) & 1.01 \\
\hline
\end{tabular}
}
\caption{Results from fits with \Eq{eq:stinglfit} ($n=1$)
for $D_L$ (\LHS) and $D_T$ (\RHS) corresponding to the Monte Carlo data 
shown in \Fig{fig:templongtrans} ($\beta=6.337, N_{\sigma}=48$). 
The fit range is $[0.6 : 8.0]~\mathrm{GeV}$.
The values in parentheses provide the fit errors. The boldface printed 
$b$-values indicate that they are fixed to zero.}
\label{fit:DallT}
\end{table*}
%-------------------------------------------------------------------------------

\begin{figure*}[tb]
 \centering
 \mbox{
 \includegraphics[angle=0,scale=0.7]{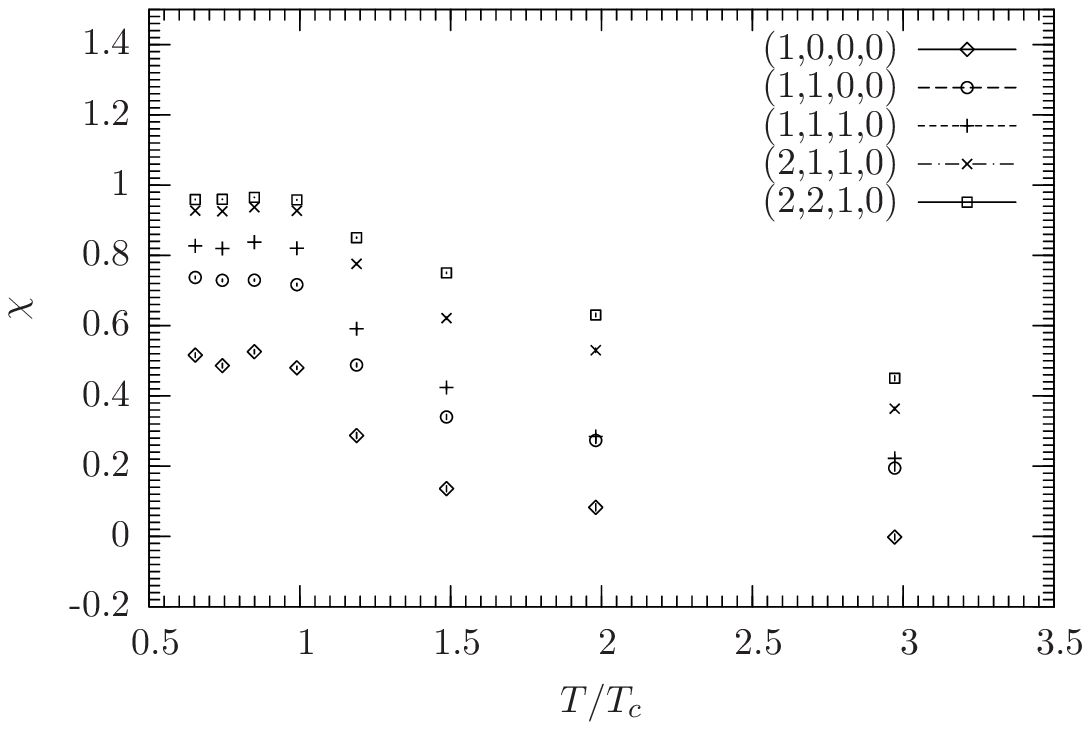} 
 \hspace*{-0.5cm}
 \includegraphics[angle=0,scale=0.7]{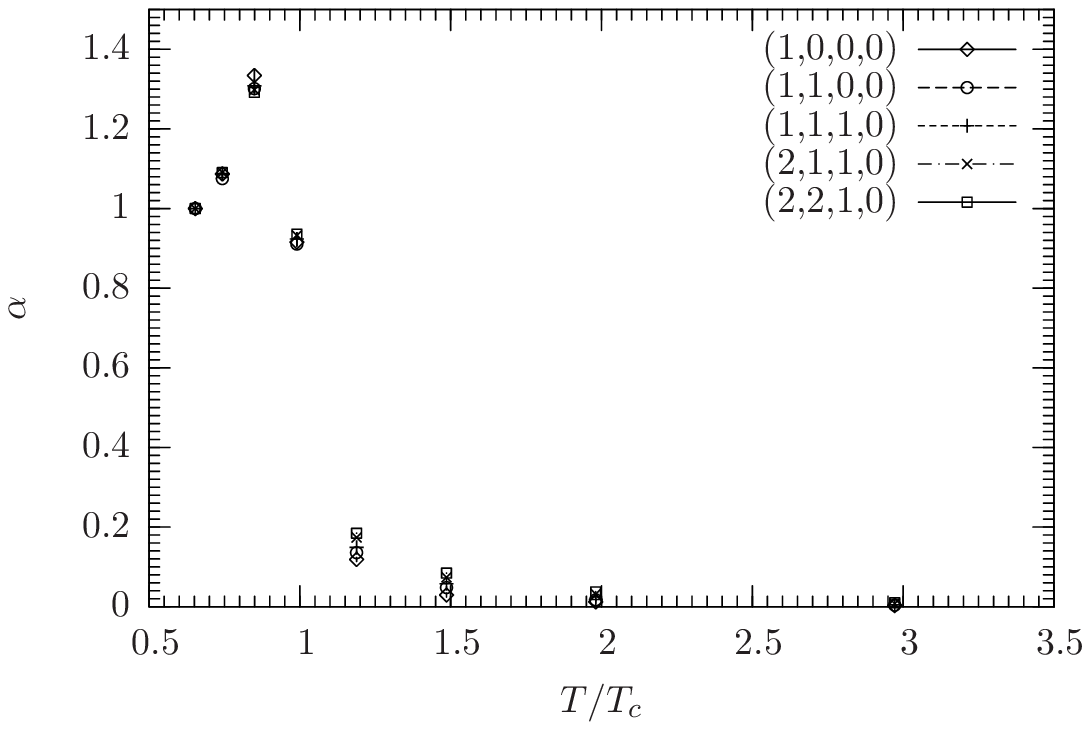} 
 }
\caption{Temperature behavior of the ratios $\chi$ (\Eq{equ:chi}, left panel) 
and $\alpha$ (\Eq{equ:alpha}, right panel) at low momenta as given in the 
legend, for a spatial lattice size $N_{\sigma}=48$ and $\beta=6.337$.}
\label{fig:ord}
\end{figure*}

\vspace{2mm}
We have tried to form quantities constructed from the 
gluon propagator which can serve as indicators
for the deconfinement transition.
First, we plot the ratio
\beq
 \chi=[D_{L}(0,T)-D_{L}(q,T)]/D_{L}(0,T)
\label{equ:chi}
\eeq
as a function of $~T/T_{c}~$ in the left panel of \Fig{fig:ord}.
%------------------------------------------------------------------------------
%------------------------------------------------------------------------------
We observe that all the curves labelled by the momentum 4-tuples in the legend 
show approximate plateaux below $T_{c}$. 
Then, passing the phase transition they suddenly fall off with slopes becoming 
slightly smaller with increasing momentum, but still with visible temperature 
sensitivity. This means that $\chi$ can be used as an indicator for the 
deconfinement transition and, moreover, the transition can be traced even at 
rather high momentum. This was not so clear from the \LHS~of
\Fig{fig:templongtranslow}, where the behavior of $D_L$ at higher momenta 
looks rather smooth.

From the behavior of $\chi$, at least in the interval 
$~0.65 \,T_c \,\aleq\, T \,\aleq\, T_c$
and at low momentum, one can conjecture the factorization
\beq
D_L(q;T) \simeq A(q) \cdot B(T)\,.
\label{equ:factor}
\eeq
Then, as long as the temperature $T$ varies in the given interval,
the change of $D_L$ can be described by a momentum independent rescaling. 
This is a rather nontrivial property from which further conclusions can be drawn.
For example, in the interpolation formula (\ref{eq:stinglfit}) above, we 
should find the mass parameter $r^2$ and the parameter $d$ to be 
(approximately) temperature independent as long as $T < T_c$. 

From the left panel of \Tab{fit:DallT} one can see that this is true
for the parameter $d$ which varies within error bars.
The variation of parameter $r^2$ is up to 30\%. 
This comparatively large variation might be explained by the fact that 
the propagators were fitted over a wide range of momenta, specifically 
from $0.6$ to $8\,\textrm{GeV}$, while the factorization we expect 
to hold only at low momenta. Indeed, 
a fit in the range up to $2.5$ GeV which includes also the zero momentum
value has shown $r^2$ to become approximately constant. It remains to be 
seen, whether this behavior in the infrared region survives the continuum 
and the thermodynamic limites which goes beyond the scope of this paper.

Let us consider another ratio 
\beq
\alpha=\dfrac{D_{L}(0,T)-D_{L}(q,T)}{D_{L}(0,T_{min})-D_{L}(q,T_{min})}\,, \quad
T_{min} = 0.65 \,T_c\,,
\label{equ:alpha}
\eeq
which according to the factorization (\ref{equ:factor}) should be approximately 
momentum independent. Indeed, this can be seen from the right panel of \Fig{fig:ord}. 
Moreover, $\alpha(q,T)$ should resemble qualitatively the temperature dependence of 
$D_L$ at $q=0$. Close to
$T_{c}$, however, $\alpha$ falls off reaching very small values at higher 
temperatures (around $2\,T_{c}$). Therefore, we conclude that both quantities 
$\chi$ (ceasing to be constant) and $\alpha$ (with its strong fall off) signal   
the finite-temperature transition. It remains to be seen, whether they also map 
out the (pseudo)critical behavior in unquenched QCD.

Let us note that our volumes are not large enough to study the infrared 
asymptotic behavior. Moreover, at the lowest momenta we expect systematic 
deviations due to finite-size effects, lattice artifacts, and Gribov 
copy effects. This concerns also the parameters $\chi$ and $\alpha$ because 
of their dependence on the value $D_L(q=0)$. The systematic 
effects will be discussed to some extent in Sections
\ref{sec:finitevolume} to \ref{sec:scaling}, in order to identify the 
momentum range, where they play only a negligible r\^ole.

Summarizing, in agreement with findings in other recent investigations 
\cite{Cucchieri:2007ta,Maas:2009fg,Fischer:2010fx,Bornyakov:2011jm,
Cucchieri:2011di} we observe the strongest response to the phase 
transition to occur in the gluonic chromoelectric sector (the longitudinal 
propagator) rather than in the gluonic chromomagnetic one (the transverse 
propagator).

\begin{figure*}[tb]
 \centering
 \mbox{
 \includegraphics[angle=0,scale=0.75]{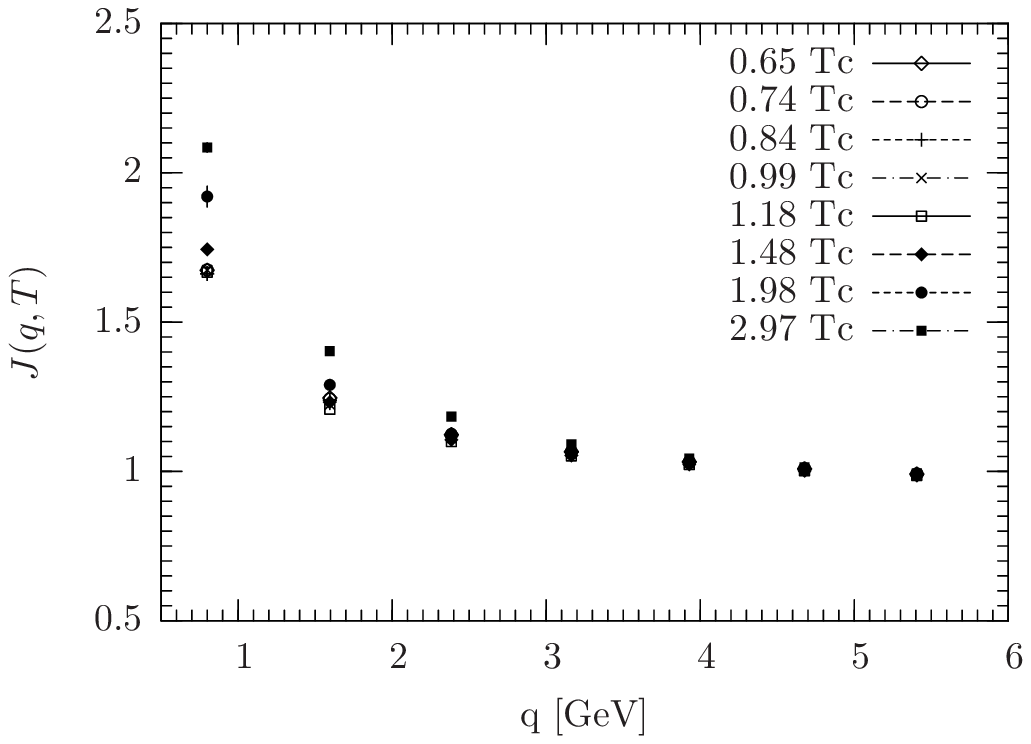} 
 \hspace*{-0.5cm}
 \includegraphics[angle=0,width=0.50\textwidth,height=0.32\textwidth]{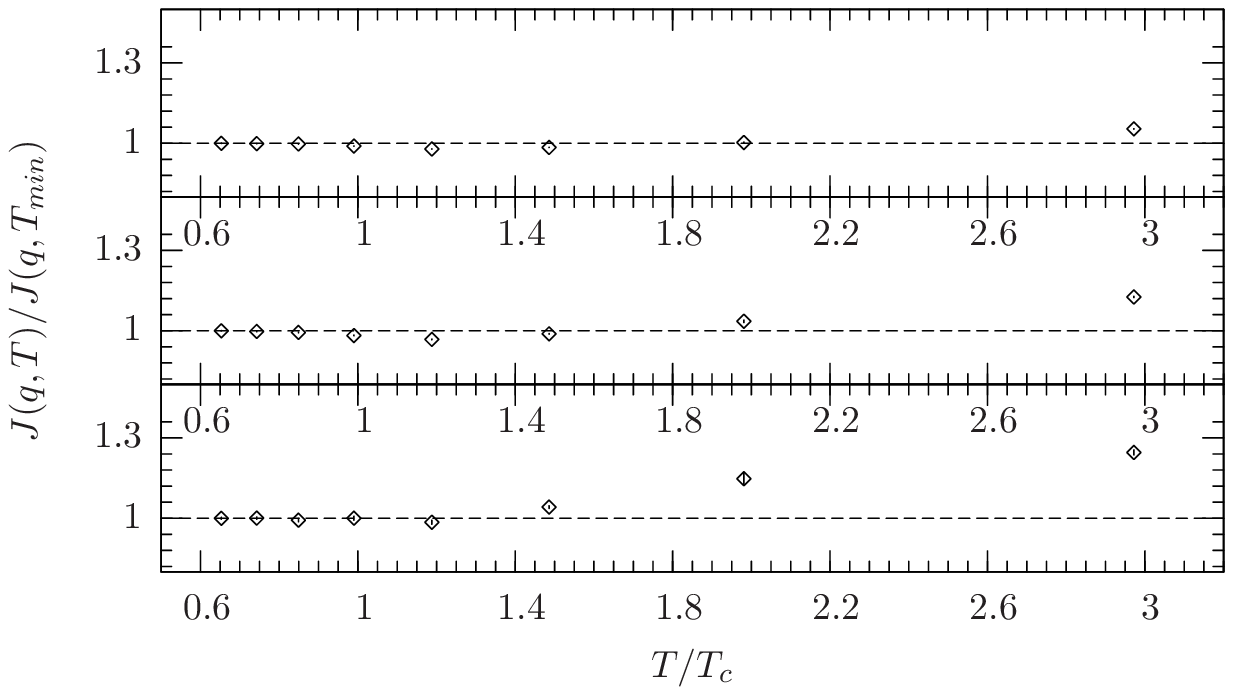} 
 }
\caption{The renormalized ghost dressing function $J(q,T)$ for various temperature
values (\LHS) and its dependence on the temperature shown for the fixed  
diagonal 3-momenta ($(k,k,k,0), k=1,2,3$) and normalized with 
$J(q,T_{min})$ for $T_{min}=0.65 T_c$ (\RHS). The lowest panel shows the lowest momentum. 
All data are obtained at $\beta=6.337$ on a lattice with spatial size $N_{\sigma}=48$. 
}
\label{fig:ghost}
\end{figure*}

We have also computed the ghost propagator according to \Eq{eq:ghost}, 
restricting it for simplicity to the diagonal three-momenta and vanishing
Matsubara frequency, $k_\mu = (k,k,k,0)$ with $k= 1, \ldots, 7$. The data are 
again normalized at $\mu=5~\mathrm{GeV}$, such that the ghost dressing function 
equals unity at $q=\mu$. The result for the latter function is displayed in 
\Fig{fig:ghost}. In comparison with the gluon propagator we see the ghost propagator 
to change relatively weakly with the temperature.\footnote{Note the use of a 
linear scale at the vertical axis in \Fig{fig:ghost}a in contrast to the logarithmic
one in \Fig{fig:templongtrans}.} This is in agreement with the 
observation in \cite{Cucchieri:2007ta}. 
An increase becomes visible at temperature values $T > 1.4 T_c$ for the
lowest momenta studied (see \Fig{fig:ghost}b). The relative insensitivity 
with respect to the temperature is the reason why we will not further 
consider the ghost propagator in what follows.

%-----------------------------------------------------------------------------
%------------------------------------------------------------------------------

\section{Finite-volume effects}
\label{sec:finitevolume}
%------------------------------

%-------------------------------------------------------------------------------
\begin{figure*}[tb]
\mbox{\hspace*{-0.2cm}
 \includegraphics[angle=0,scale=0.7]{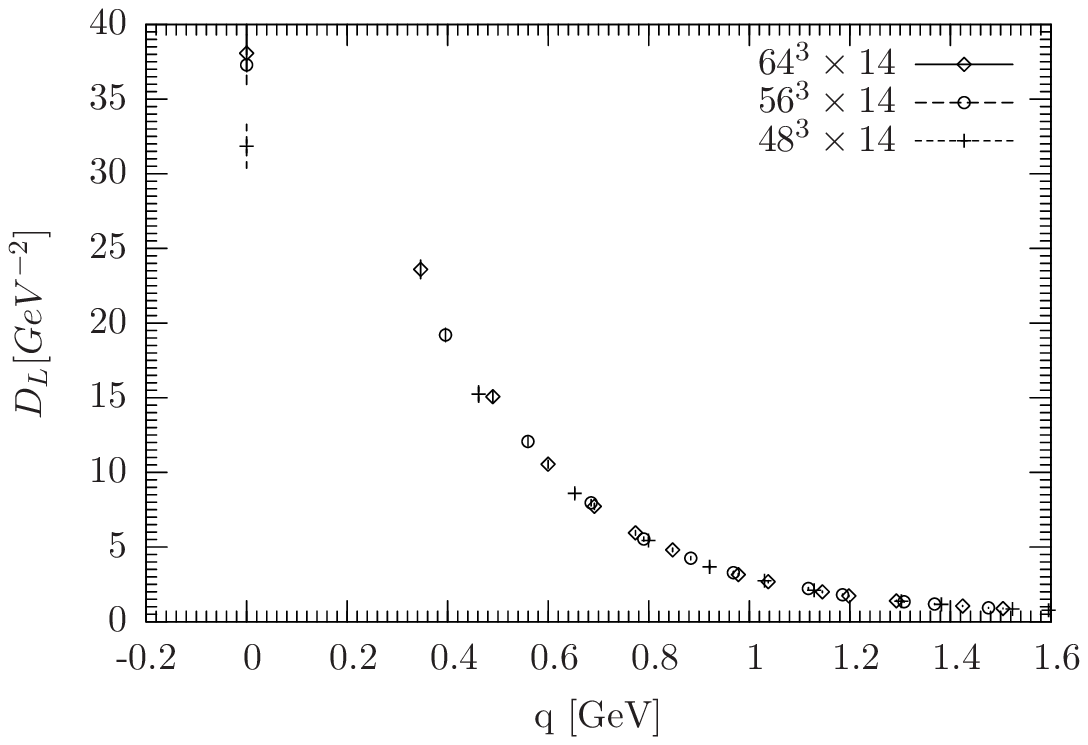} 
 \hspace*{-1.2cm}
 \includegraphics[angle=0,scale=0.7]{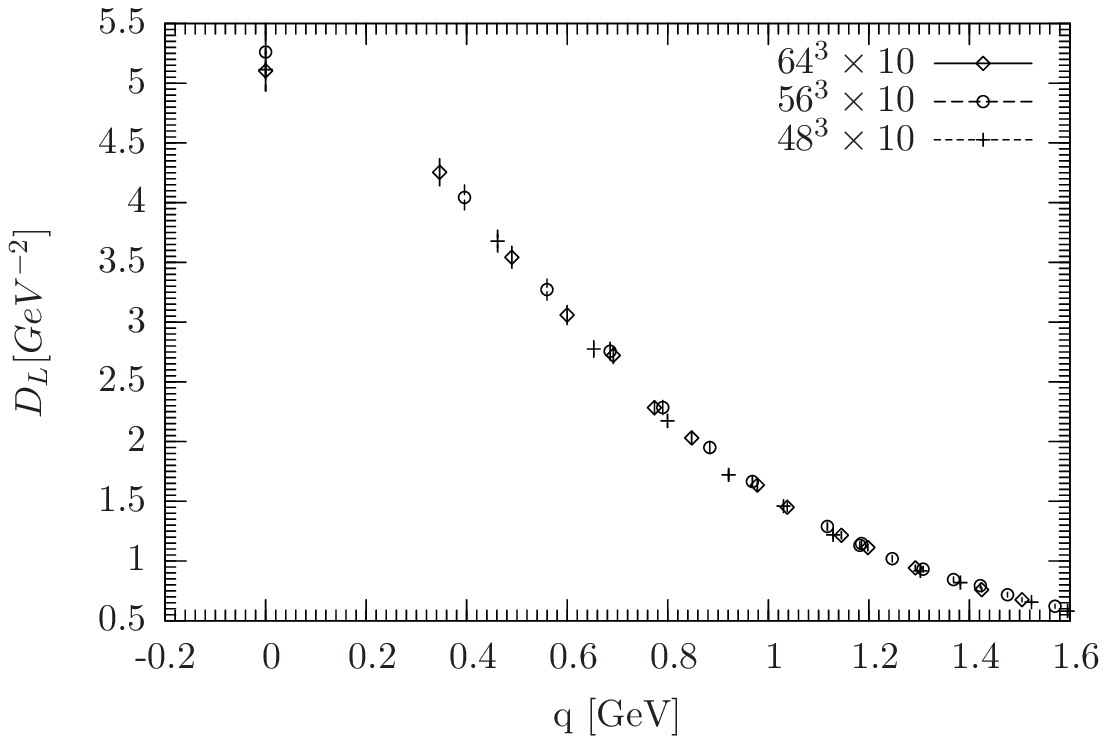} 
 }
\caption{Finite-size effect study for $D_L$ at $\beta=6.337$. 
\LHS: ~$T=0.86\,T_c$, \RHS: ~$T=1.20\,T_c$. 
}
\label{fig:volumelong}
\bigskip
\mbox{\hspace*{-0.1cm}
 \includegraphics[angle=0,scale=0.7]{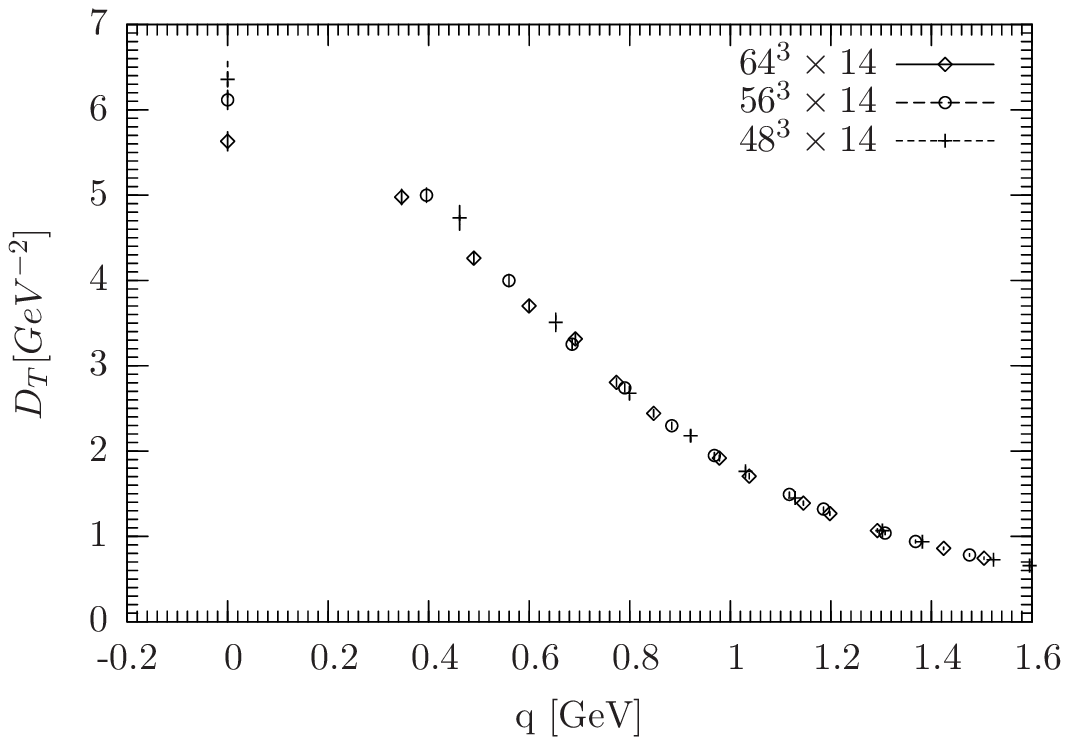} 
 \hspace*{-1.2cm}
 \includegraphics[angle=0,scale=0.7]{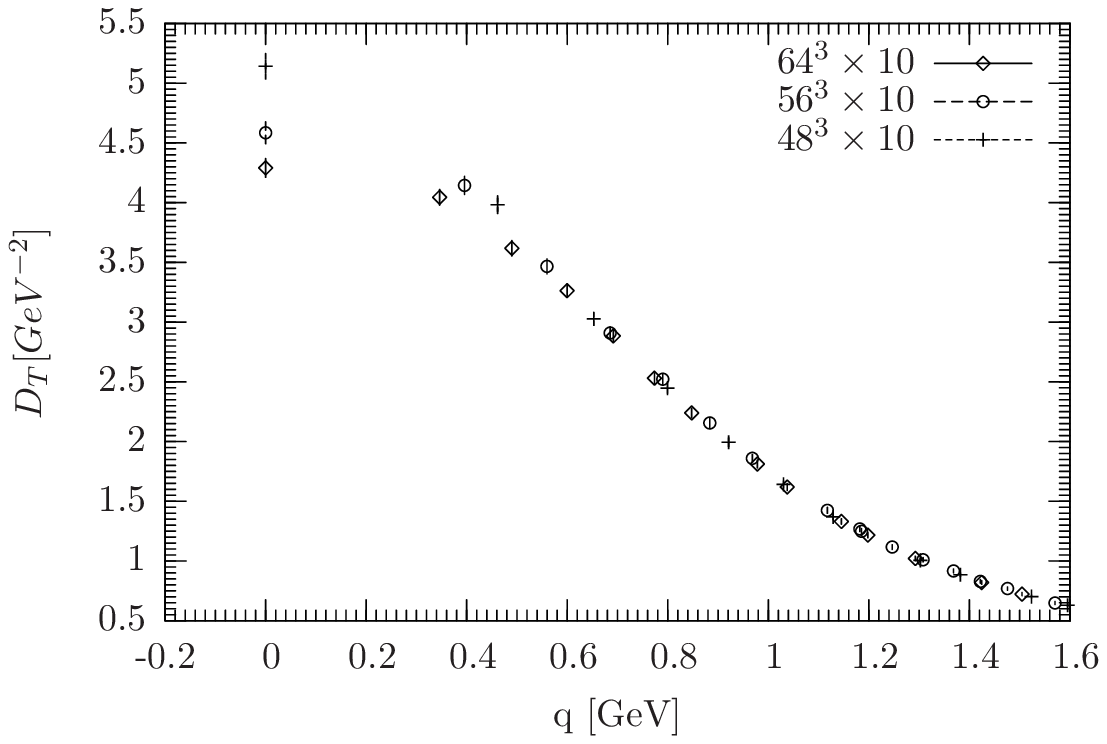} 
 }
\caption{Same as in \Fig{fig:volumelong} but for $D_T$.}
\label{fig:volumetrans}
\end{figure*}
%---------------------------------------------------------------------------
In order to estimate finite-volume effects we compare the data shown before 
with data obtained on even larger spatial volumes while keeping fixed 
the coupling (at $\beta=6.337$) and two temperature values, 
$T=0.86\,T_{c}$ (confinement) and $T=1.2\,T_{c}$ (deconfinement), 
respectively. The linear spatial extent varies from 
$48 a = 2.64 \,\mathrm{fm}$ to $64 a= 3.52 \,\mathrm{fm}$ 
(see also the middle section in \Tab{tab:numbers}).

In Figs. \ref{fig:volumelong} and \ref{fig:volumetrans} we show the 
corresponding plots for $D_L$ and $D_T$, respectively. In all four cases we 
observe the effects to be small for momenta above $0.6~\mathrm{GeV}$.% 
\footnote{
Below $T_c$ the transverse propagator changes by less than 12~\%, the longitudinal 
one by less than 5~\%. Above $T_c$ the transverse propagator varies by less 
than 8~\% and the longitudinal one by less than 11~\%.} 
For lower momenta, especially at zero momentum, systematic deviations become
more visible. With increasing volume the infrared values of $D_L$ seem 
to rise, whereas for $D_T$ the opposite is the case.
This behavior has already been reported for pure gauge theories 
in \cite{Cucchieri:2007md,Bornyakov:2010nc} for $SU(2)$ and in \cite{Bornyakov:2011jm}  
for $SU(3)$, respectively.

\section{Gribov copy effects}
\label{sec:gribov}
%----------------------------

In order to study Gribov copy effects we compare ``first'', i.e. randomly 
occuring copies (\fc) with ``best'' copies (\bc). 
The latter were produced as follows.

We searched for copies within all $3^3=27\,$ ${\bf Z}(3)$ sectors
characterized by the phase of the spatial Polyakov loops, i.e.
Polyakov loops in one of the three spatial directions.
For this purpose the ${\bf Z}(3)$ flipping operations
\cite{Bogolubsky:2005wf,Bornyakov:2011jm} were carried out on all
link variables $U_{x,i}$ ($i=1,2,3$) attached and orthogonal to a
3d hyperplane with fixed $x_i$ by
multiplying them with $\exp{\{\pm 2\pi i/3\}}$.  Such global flips are
equivalent to non-periodic gauge transformations and do not change the
pure gauge action.  For the $4$th direction, we stick to the sector with
$ |arg~P| < \pi/3$ which provides maximal values of the functional
(\ref{eq:functional}) at the $\beta$-values considered in this 
section~\cite{Bornyakov:2011jm}. Thus, the flip operations combine
for each lattice field configuration the $27$ distinct gauge orbits
of strictly periodic gauge transformations into one larger gauge orbit.

The number of copies actually considered in each of the 27 sectors
depends on the rate of convergence (with increasing number of 
investigated copies) of the propagator values assigned to the best copy, 
in particular at zero momentum. 
From our experience with $SU(3)$ theory \cite{Bornyakov:2011jm}  we expect
that the effect of considering gauge copies in different flip-sectors is
more important than probing additional gauge copies in each sector. For
this reason and to save CPU time we have considered one gauge copy for
every ${\bf Z}(3)$-sector; therefore, in total $n_{copy}=27$ gauge
copies for every configuration.

To each copy the simulated annealing algorithm with consecutive overrelaxation
was applied in order to fix the gauge. We take the copy with maximal value 
of the functional (\ref{eq:functional}) as our best realization of the 
global maximum and denote it as best (``\bc'') copy. 

The parameters of the SA algorithm in the study of Gribov copies were 
slightly different from those described above in \Sec{sec:propagators}.  
2000 SA combined simulation sweeps with a ratio 11:1 between microcanonical 
and heat bath sweeps were applied starting with  
$T_{sa}=0.5$ and ending at $T_{sa}=0.0033$.

Since this procedure is quite CPU time consuming we restricted this investigation
to coarser lattices $6\times 28^3$ and $8\times 28^3$ with larger lattice spacing,
such that both the temperature values $T=0.86 \,T_c$ and $T=1.20 \,T_c$, respectively,
as well as the physical 3d volume $(2.64 \,\mathrm{fm})^3$ were approximately 
reproduced.

%-----------------------------------------------------------------------------
\begin{figure*}[tb]
 \mbox{\hspace*{-0.2cm}
 \includegraphics[angle=0,scale=0.7]{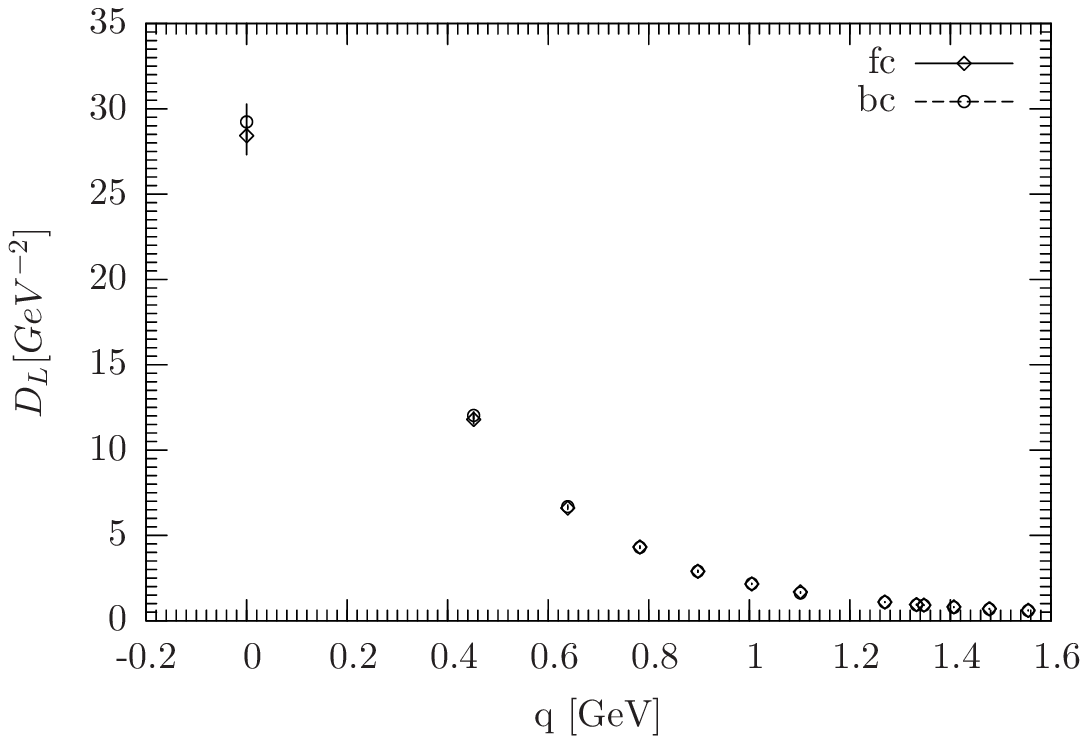} 
 \hspace*{-1.2cm}
 \includegraphics[angle=0,scale=0.7]{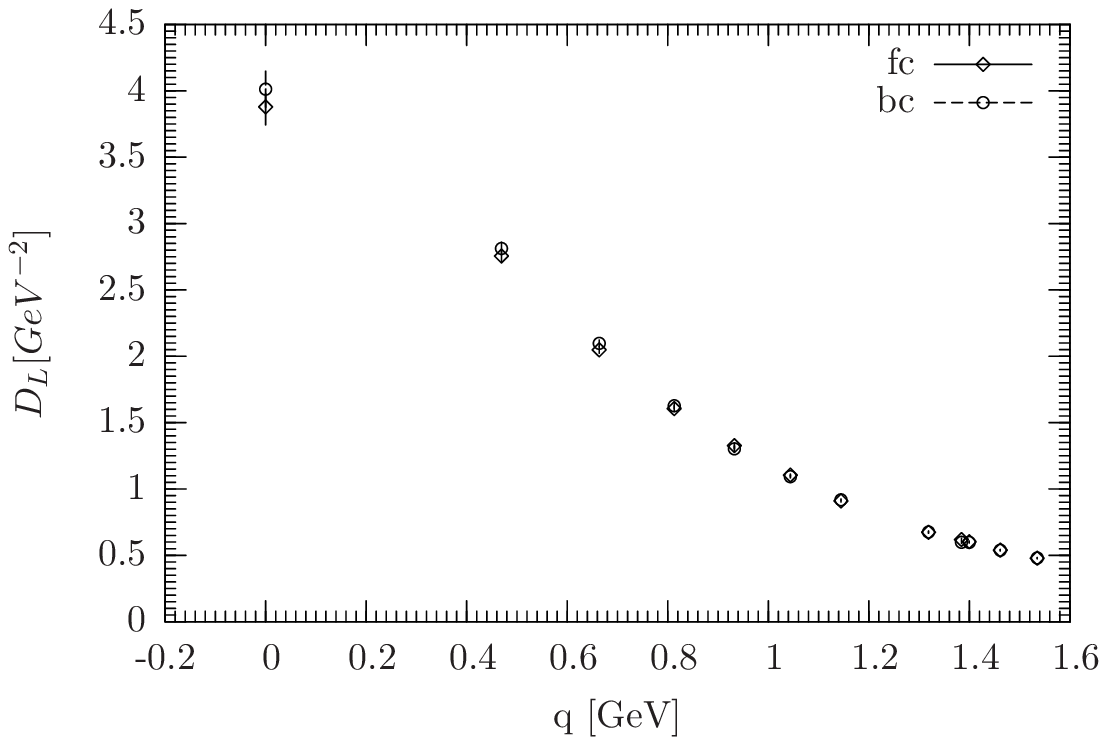} 
 }
\caption{Comparison of the \bc~ with the \fc~ Gribov copy result
for the longitudinal propagator $D_L$ (unrenormalized)  
(\LHS: $T=0.86\,T_c$, \RHS: $T=1.20\,T_c$).} 
\label{fig:gribov_long}
%------------------------------------------------------------------------------
\bigskip
 \mbox{\hspace*{-0.2cm}
 \includegraphics[angle=0,scale=0.7]{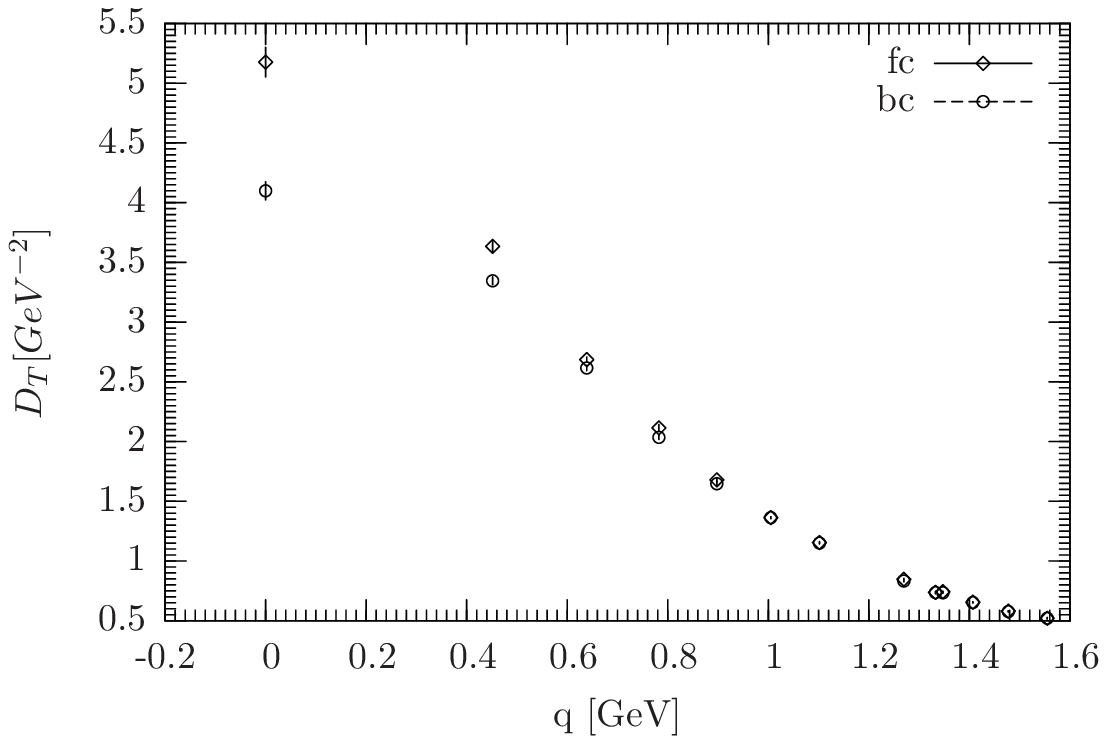} 
\hspace*{-1.2cm}
 \includegraphics[angle=0,scale=0.7]{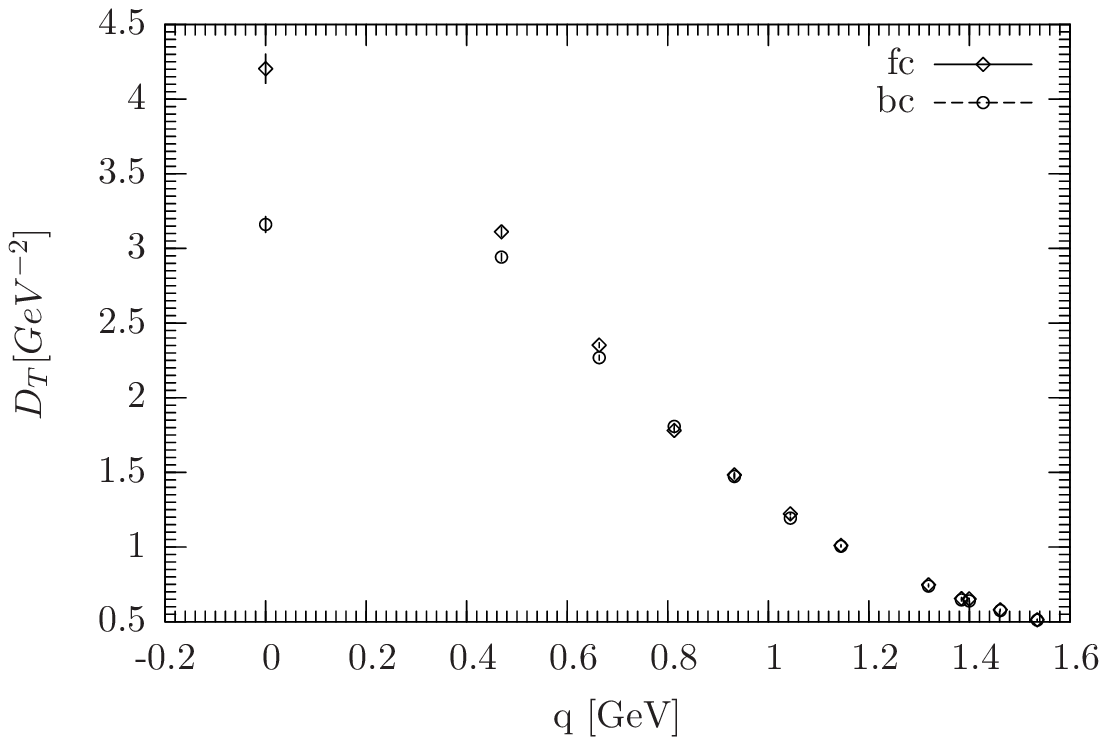} 
 }
\caption{Same as in \Fig{fig:gribov_long} but for the transverse
propagator $D_T$ (\LHS: ~$T=0.86\, T_c$, \RHS: ~$T=1.20\,T_c$).
}
\label{fig:gribov_trans}
\end{figure*}
%------------------------------------------------------------------------------

In Figs. \ref{fig:gribov_long} and \ref{fig:gribov_trans} we compare \bc~
with \fc~ results for the gluon propagators $D_L$ and $D_T$, respectively.
As one can see, $D_L$ is almost insensitive to the choice of Gribov copies
(at least, for the comparatively small values of $N_{\tau}$ we consider),
as has been already reported in \cite{Bornyakov:2010nc} for the $SU(2)$
case and in \cite{Bornyakov:2011jm} for the $SU(3)$ case.
On the contrary, the transverse propagator is strongly affected in the
infrared region. This observation is independent of the temperature.
Moreover, we see that the transverse gluon propagator values in
the infrared become lowered for \bc~  compared with \fc~ results.
These observations resemble those made already in
\cite{Bornyakov:2010nc} and \cite{Bornyakov:2011jm}.

\vspace{1mm}
The main conclusion of this section is that Gribov copy effects may be
neglected for all nonzero momenta in the case of the longitudial
propagator (at least, for comparatively small values of $N_{\tau}$),
and for momenta above 800 MeV in the case of the transverse propagator.
The momentum range where the last statement is true might depend on the
temperature.

\section{Scaling and continuum limit}
\label{sec:scaling}
%------------------------------------

\begin{figure*}[tb]
\mbox{
 \includegraphics[angle=0,scale=0.7]{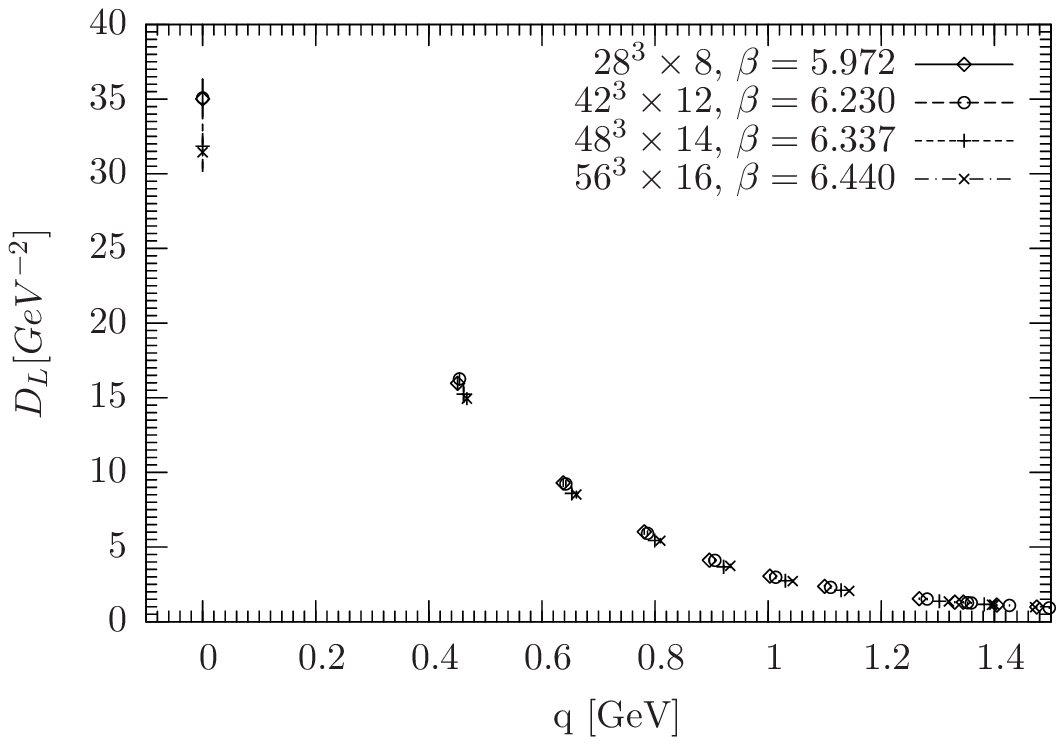} 
\hspace*{-1.1cm}
 \includegraphics[angle=0,scale=0.7]{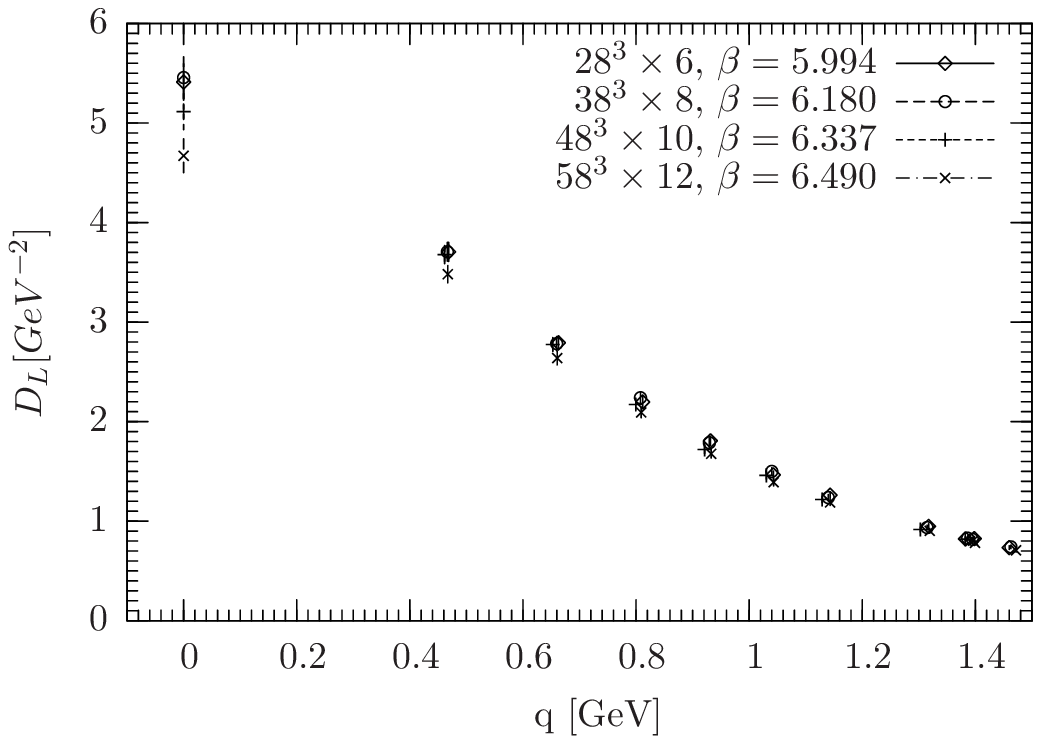} 
 }
\caption{The longitudinal propagator $D_L$, renormalized at $\mu=5~\mathrm{GeV}$, 
obtained for fixed physical volume and temperature but varying $a=a(\beta)$. 
\LHS: ~$T=0.86\, T_c$, \RHS: ~$T=1.20\,T_c$.
}
\label{fig:scalelong}
\bigskip
\mbox{
 \includegraphics[angle=0,scale=0.7]{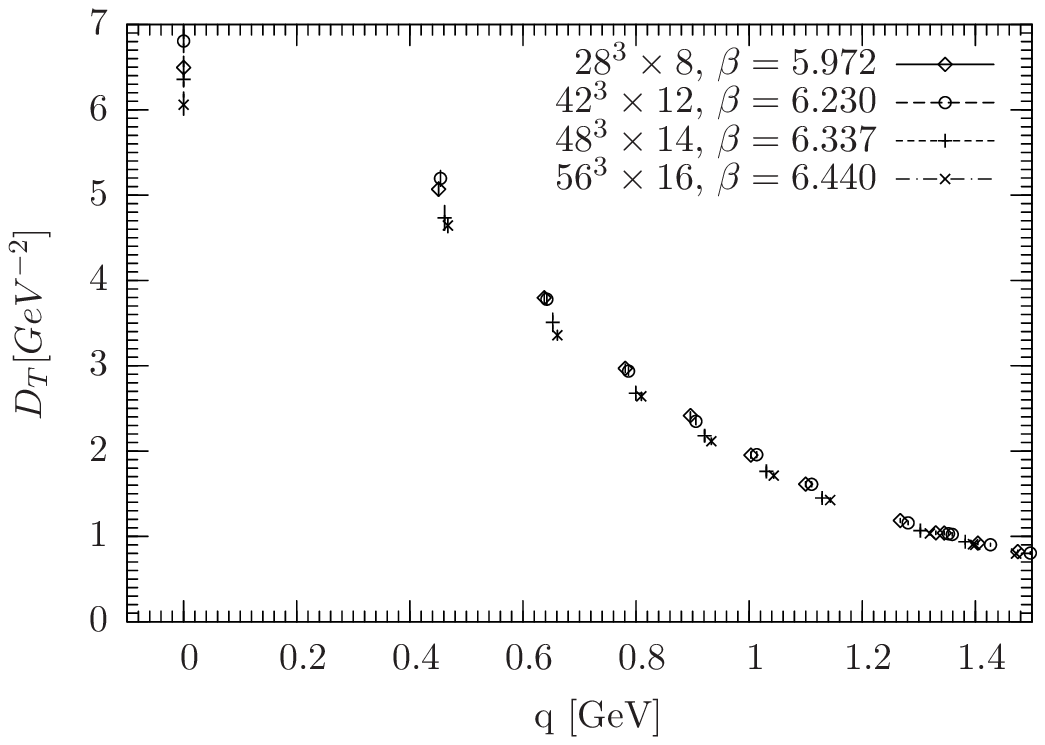} 
\hspace*{-1cm}
 \includegraphics[angle=0,scale=0.7]{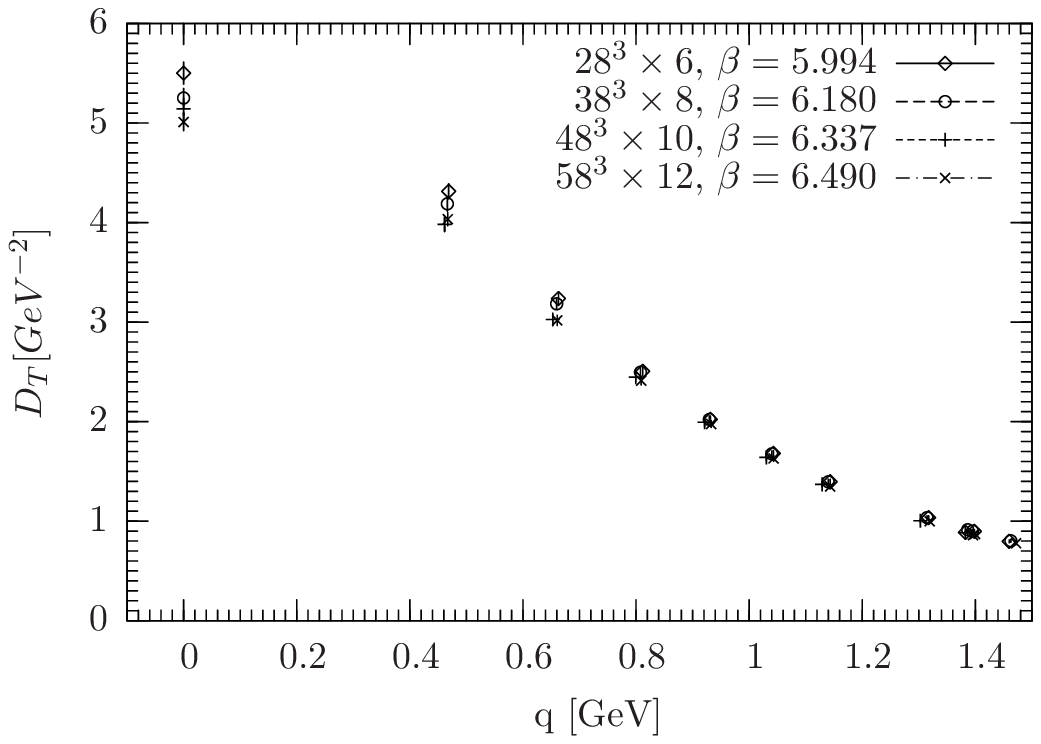} 
 }
\caption{
 Same as in \Fig{fig:scalelong} but for the transverse propagator $D_T$.
\LHS: ~$T=0.86\, T_c$, \RHS: ~$T=1.20\,T_c$}
\label{fig:scaletrans}
\end{figure*}

In order to check for good scaling properties we have used the same reference 
values for the temperature below and above $T_{c}$ as discussed before 
(i.e., $0.86\,T_{c}$ and $1.20\,T_{c}$).
We kept also the spatial volume fixed at $(2.7~\mathrm{fm})^3$. 
We compare the renormalized propagators at four different values for 
the lattice spacing $a(\beta)$ (see \Tab{tab:numbers}). Our results 
are displayed for the momentum range up to 1.5 GeV 
in \Fig{fig:scalelong} for $D_L$ and in \Fig{fig:scaletrans} for $D_T$, 
respectively. Gauge fixing has been carried out as originally described in 
\Sec{sec:propagators}.

We provide the renormalization factors for
\beq  \label{eq:renorm}
D_{L,T}(q,\mu) \equiv Z_{L,T}(a,\mu)~D^{bare}_{L,T}(q,a) 
\eeq
in the left panel of \Tab{tab:Zfactors-fitinterpo}.
As expected the $Z$-factors of $D_L$ and $D_T$ approximately
agree.

From Figs. \ref{fig:scalelong} and \ref{fig:scaletrans} we
see that the scaling violations happen to be reasonably small for 
momenta above $0.8~\mathrm{GeV}$. This shows that our choice of 
$a=0.055~\mathrm{fm}$ for $\beta=6.337$ was already close to the 
continuum limit. 

%------------------------------------------------------------------------------
%------------------------------------------------------------------------------

In order to study the $a$-dependence at five particular physical momenta $p$ 
we need interpolations of the momentum dependence in between the data points. 
For the fit within the interval $0.6 \,\mathrm{GeV} \le q \le 3.0 \,\mathrm{GeV}$ 
we have used again \Eq{eq:stinglfit} with parameter $b$ fixed to zero. 
The values of the fit parameters are displayed in the right hand panels of
\Tab{tab:Zfactors-fitinterpo}. In all cases we find $\chi^{2}$-values per 
degree of freedom around or below unity. 

%------------------------------------------------------------------------------
\begin{table*}
%------------------------------------------------------------------------------

\mbox{
\setlength{\tabcolsep}{4.0pt}
\begin{tabular}{|c|c|c|c|c|c|}
\hline
\multicolumn{4}{|c|}{Parameters} & \multicolumn{2}{|c|}{$Z$-factors} \\
\hline
$T/T_{c}$ & $\beta$ & $N_{\sigma}$ & $N_{\tau}$ & $Z_{T}$ 
                                                & $Z_{L}$ \\
\hline
0.86 & 5.972 & 28 &  8 & 1.43 & 1.43 \\
0.86 & 6.230 & 42 & 12 & 1.45 & 1.47 \\
0.86 & 6.337 & 48 & 14 & 1.48 & 1.53 \\
0.86 & 6.440 & 56 & 16 & 1.64 & 1.66 \\
1.20 & 5.994 & 28 &  6 & 1.46 & 1.46 \\
1.20 & 6.180 & 38 &  8 & 1.52 & 1.52 \\
1.20 & 6.337 & 48 & 10 & 1.62 & 1.63 \\
1.20 & 6.490 & 58 & 12 & 1.62 & 1.65 \\
\hline
\end{tabular}
}
%------------------------------------------------------------------------------
\mbox{
\setlength{\tabcolsep}{1.0pt}
\begin{tabular}{|c|c|c|c|}
\hline
\multicolumn{4}{|c|}{$D_L$ fits} \\
\hline
$~r^2(\mathrm{GeV}^{2})~$ & $d(\mathrm{GeV}^{-2})$ & $c(\mathrm{GeV}^{2})$ 
                                                   & $\chi_{df}^{2}$ \\
\hline
 0.317(20) & 0.138(24) & 4.67(26) & 0.30 \\
 0.254(9) & 0.224(7)  & 3.90(8) & 0.44 \\
 0.262(12) & 0.224(11) & 3.80(12) & 0.42 \\
 0.256(7) & 0.220(6)  & 3.86(7) & 0.24 \\
 0.995(37) & 0.153(10) & 5.46(24) & 0.80 \\
 0.985(20) & 0.163(6)  & 5.34(13) & 0.28 \\
 0.960(19) & 0.180(7)  & 4.96(13) & 0.22 \\
 1.018(18) & 0.162(5) & 5.27(11) & 0.06 \\
\hline
\end{tabular}
}
%------------------------------------------------------------------------------
\mbox{
\setlength{\tabcolsep}{1.0pt}
\begin{tabular}{|c|c|c|c|}
\hline
\multicolumn{4}{|c|}{$D_T$ fits} \\
\hline
$~r^2(\mathrm{GeV}^{2})~$ & $d(\mathrm{GeV}^{-2})$ & $c(\mathrm{GeV}^{2})$  
                                                   & $\chi_{df}^{2}$ \\
\hline
 0.810(23) & 0.148(7) & 5.49(17) & 1.19 \\
 0.835(16) & 0.151(5) & 5.69(12) & 0.52 \\
 0.867(18) & 0.142(6) & 5.62(14) & 0.14 \\
 0.880(15) & 0.143(4) & 5.65(11) & 0.36 \\
 0.894(26) & 0.144(7) & 5.55(18) & 1.10 \\
 0.924(22) & 0.142(6) & 5.71(16) & 0.57 \\
 0.982(27) & 0.133(8) & 5.87(21) & 0.59 \\
 0.963(19) & 0.140(5) & 5.77(13) & 0.45 \\
\hline
\end{tabular}
}
\caption{{\bf Left panel:} Renormalization factors $Z_{L,T}$ of the renormalized 
propagators $D_{T,L}(q, \mu)$ according to \Eq{eq:renorm}. 
The renormalization point is $~\mu=5~\mathrm{GeV}$. ~~
{\bf Right panels:} Fit parameters and $\chi_{df}^{2}$ 
for fits of $D_{L}$ (\LHS) and $D_{T}$ (\RHS) using the generic 
fit function $D(q^2)$ acc. to \Eq{eq:stinglfit}, but with $b=0$.
The fit range is restricted to $[0.6 : 3.0]~\mathrm{GeV}$.
The fit errors are indicated in parentheses. 
}
\label{tab:Zfactors-fitinterpo}
\end{table*}
%------------------------------------------------------------------------------

\begin{figure*}[tb]
 \mbox{\hspace*{-0.1cm}
 \includegraphics[angle=0,scale=0.7]{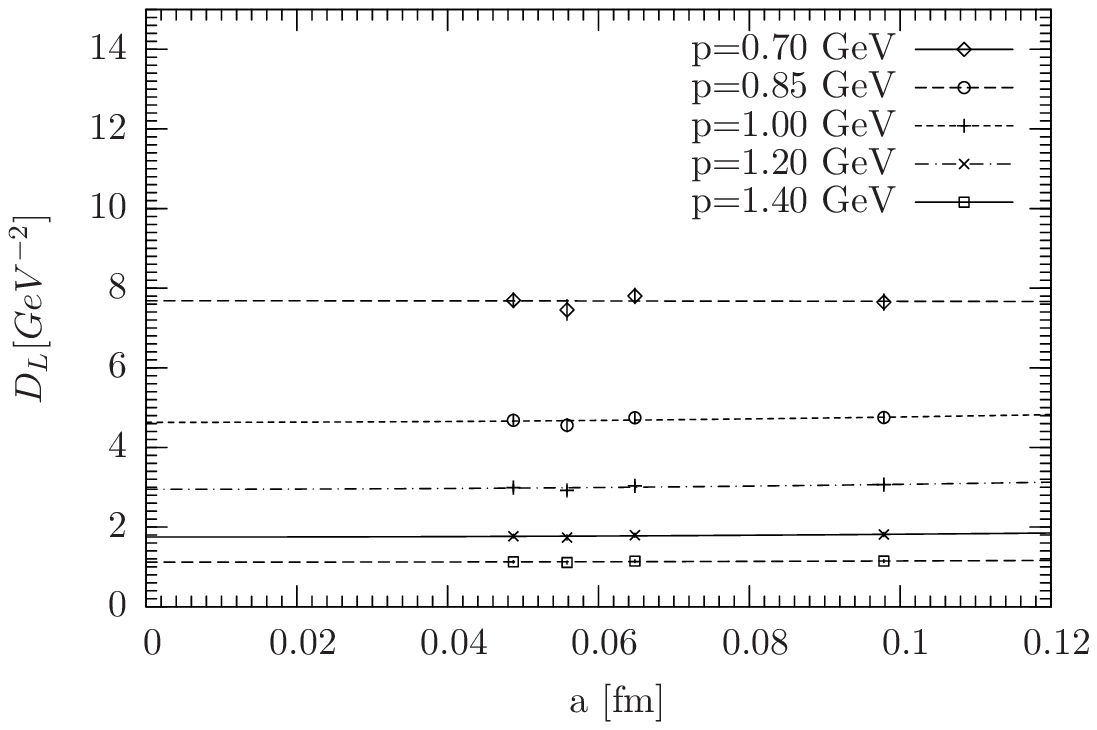}  
\hspace*{-1.3cm}
 \includegraphics[angle=0,scale=0.7]{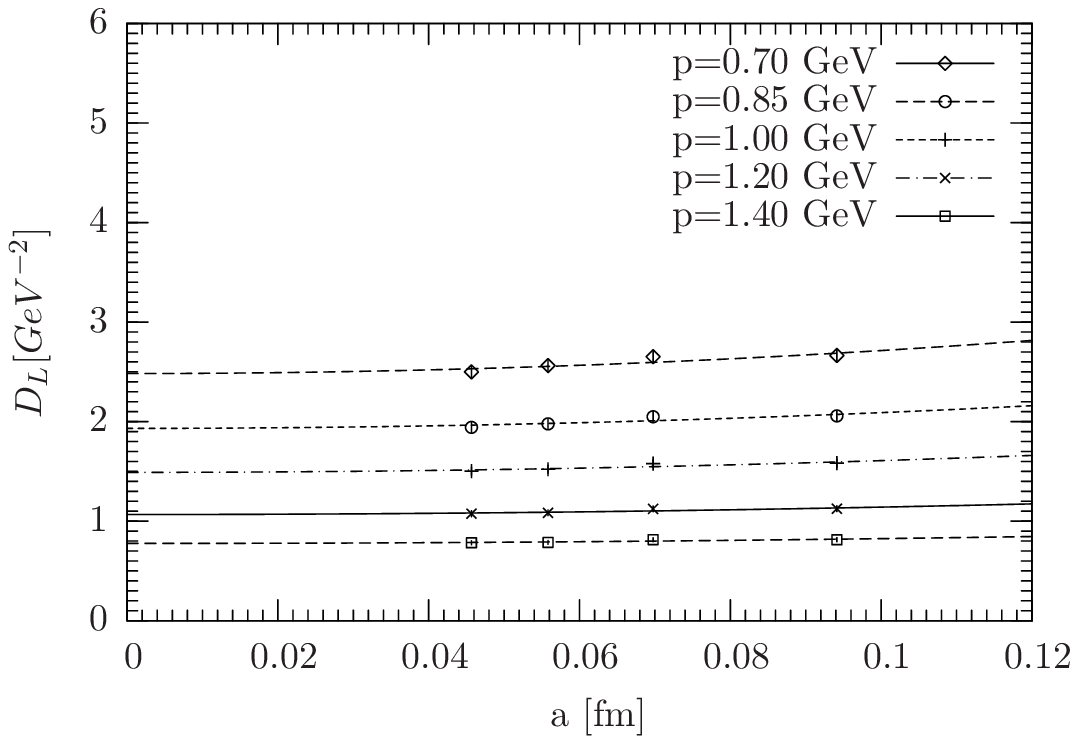} 
 }
\caption{
$D_L$ vs. lattice spacing $a$ for a set of different 
preselected momenta $p$. \LHS ~$T=0.86~T_c$; \RHS ~$T=1.20~T_c$.
}
\label{fig:fitt_interpo_dl}
%------------------------------------------------------------------------------
\bigskip
 \mbox{\hspace*{-0.1cm}
 \includegraphics[angle=0,scale=0.7]{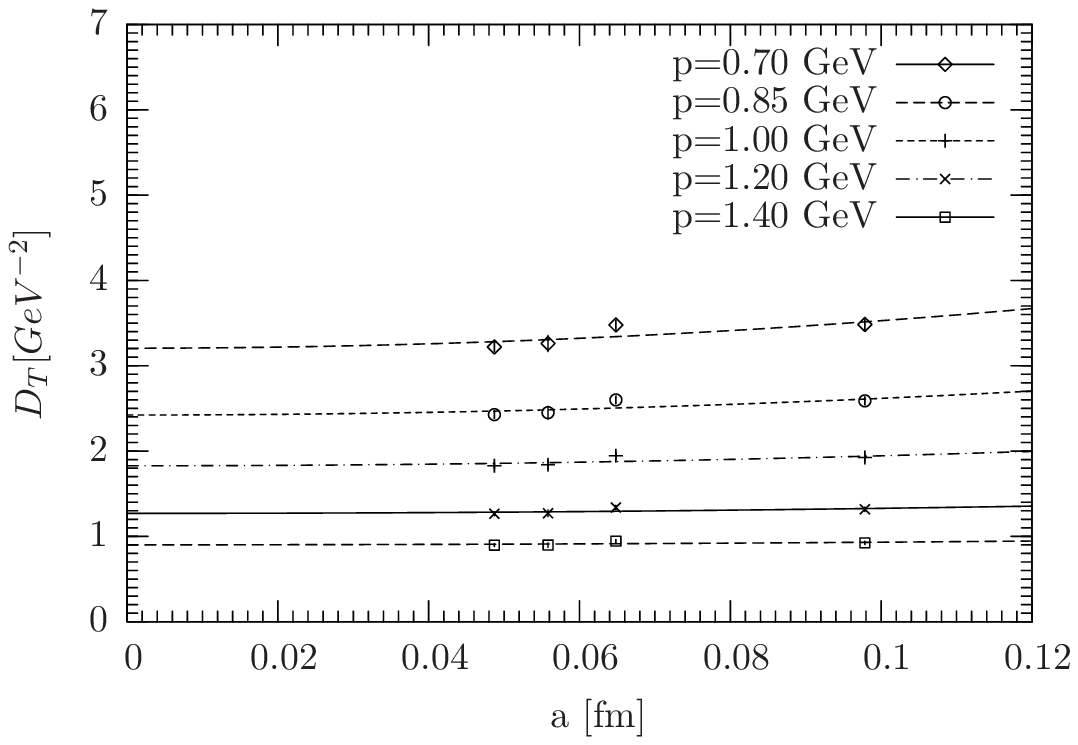} 
\hspace*{-1.2cm}
 \includegraphics[angle=0,scale=0.7]{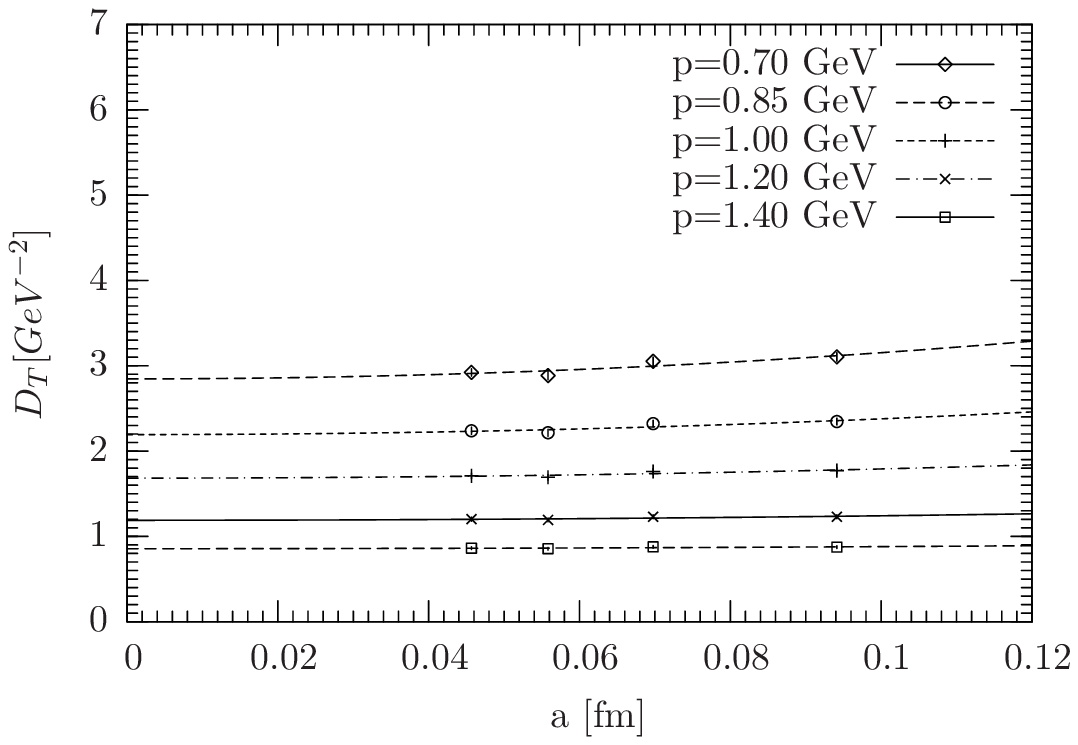} 
 }
\caption{Same as in \Fig{fig:fitt_interpo_dl} but for $D_T$.
\LHS ~$T=0.86~T_c$; \RHS ~$T=1.20~T_c$.}
\label{fig:fitt_interpo_dt}
\end{figure*}

The propagators, now interpolated to the set of selected momentum 
values, are shown in \Fig{fig:fitt_interpo_dl} for $D_L$ and in 
\Fig{fig:fitt_interpo_dt} for $D_T$, respectively, as functions 
of the lattice spacing $a$. We show them together with the respective 
fit curves 
\beq \label{eq:fit_a2}
D(a; p) =D_0 + B \cdot a^2
\eeq
assuming only $O(a^2)$ lattice artifacts. 
%------------------------------------------------------------------------------
%------------------------------------------------------------------------------
The corresponding fit results are collected in \Tab{tab:fitextrapo}. 
The respective fit parameters $D_0$ represent the continuum limit 
values of the propagators at the preselected momenta.

%------------------------------------------------------------------------------
\begin{table*}
\mbox{
\setlength{\tabcolsep}{1.0pt}
\begin{tabular}{|c|c|c|c|c|c|}
\hline
\multicolumn{2}{|c|}{Parameters} & \multicolumn{2}{|c|}{$D_L$ fits} 
                        & \multicolumn{2}{|c|}{$D_T$ fits}    \\
\hline
$~T/T_{c}~$ & $~p(GeV)~$ & $B$ & $D_0(GeV^{-2})$ & $B$ & $D_0(GeV^{-2})$ \\ 
\hline
0.86 & 0.70 & -1.3(28.1) & 7.68(16) & 32.3(20.0) & 3.20(11) \\
0.86 & 0.85 & 13.5(14.5) & 4.63(8) & 19.5(14.0) & 2.42(8) \\
0.86 & 1.00 & 12.3(7.9)  & 2.95(4) & 11.7(9.8)  & 1.83(5) \\
0.86 & 1.20 &  7.0(4.1)  & 1.75(2) &  5.9(6.4)  & 1.27(4) \\
0.86 & 1.40 &  3.0(2.6)  & 1.12(1) &  3.2(4.4)  & 0.90(2) \\
1.20 & 0.70 & 23.1(9.3)  & 2.48(5) & 30.7(11.0) & 2.84(6) \\
1.20 & 0.85 & 15.8(6.5)  & 1.93(4) & 18.5(7.4)  & 2.19(4) \\
1.20 & 1.00 & 11.8(4.7)  & 1.49(2) & 10.8(4.8)  & 1.68(2) \\
1.20 & 1.20 &  7.4(3.3)  & 1.07(2) &  5.3(3.0)  & 1.19(2) \\
1.20 & 1.40 &  4.8(2.2)  & 0.77(1) &  2.4(1.8)  & 0.86(1) \\
\hline
\end{tabular}
}
\caption{Results of the fits for $D_L$ (\LHS) and $D_{T}$ (\RHS) as a 
function of the lattice spacing $~a~$ using the fit function $D(a; p)$ 
acc. to \Eq{eq:fit_a2}. The errors of the fit parameters are given in
parentheses. $\chi_{df}^{2}$ in all cases is close or well below unity. 
See also Figs. \ref{fig:fitt_interpo_dl} and \ref{fig:fitt_interpo_dt}.}
\label{tab:fitextrapo}
\end{table*}
%------------------------------------------------------------------------------

\begin{figure*}[tb]
 \mbox{\hspace*{-0.1cm}
 \includegraphics[angle=0,scale=0.7]{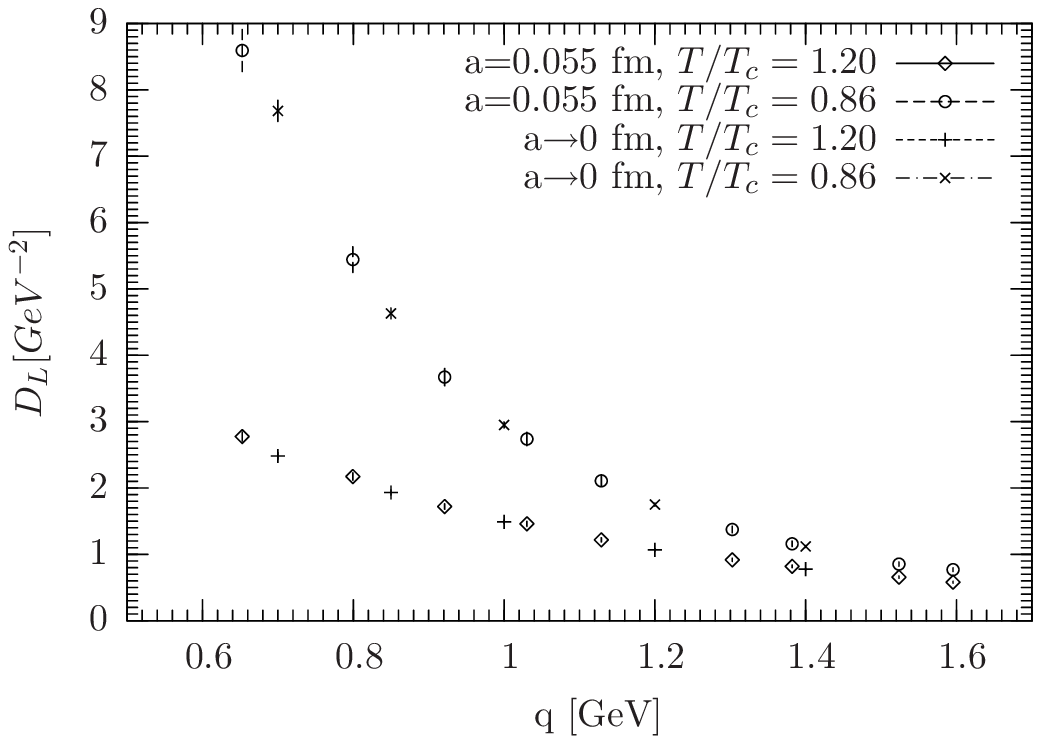} 
\hspace*{-1.5cm}
 \includegraphics[angle=0,scale=0.7]{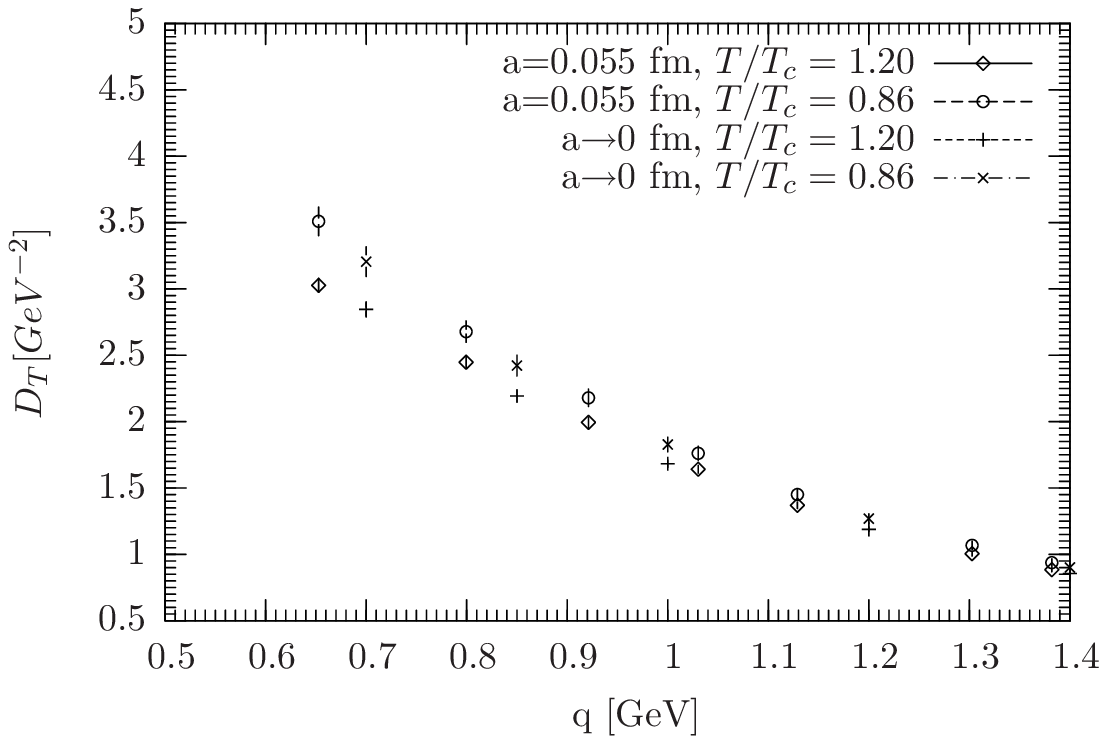} 
 }
\caption{Comparison of the renormalized propagators $D_L(q)$ (\LHS) and 
$D_T(q)$ (\RHS) obtained from the Monte Carlo simulation at $\beta=6.337$ 
with some continuum limit extrapolated values.}
\label{fig:comp_to_conti}
\end{figure*}

\begin{figure*}[tb]
 \centering
 \mbox{
 \includegraphics[angle=0,scale=0.65]{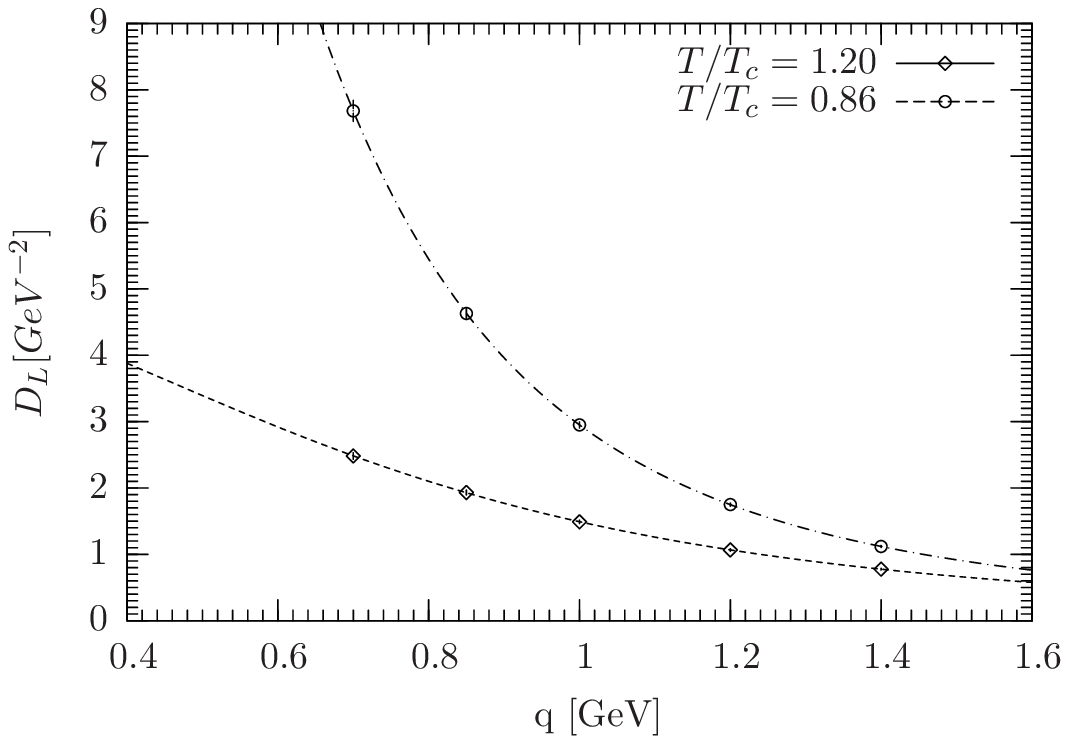} 
 \includegraphics[angle=0,scale=0.65]{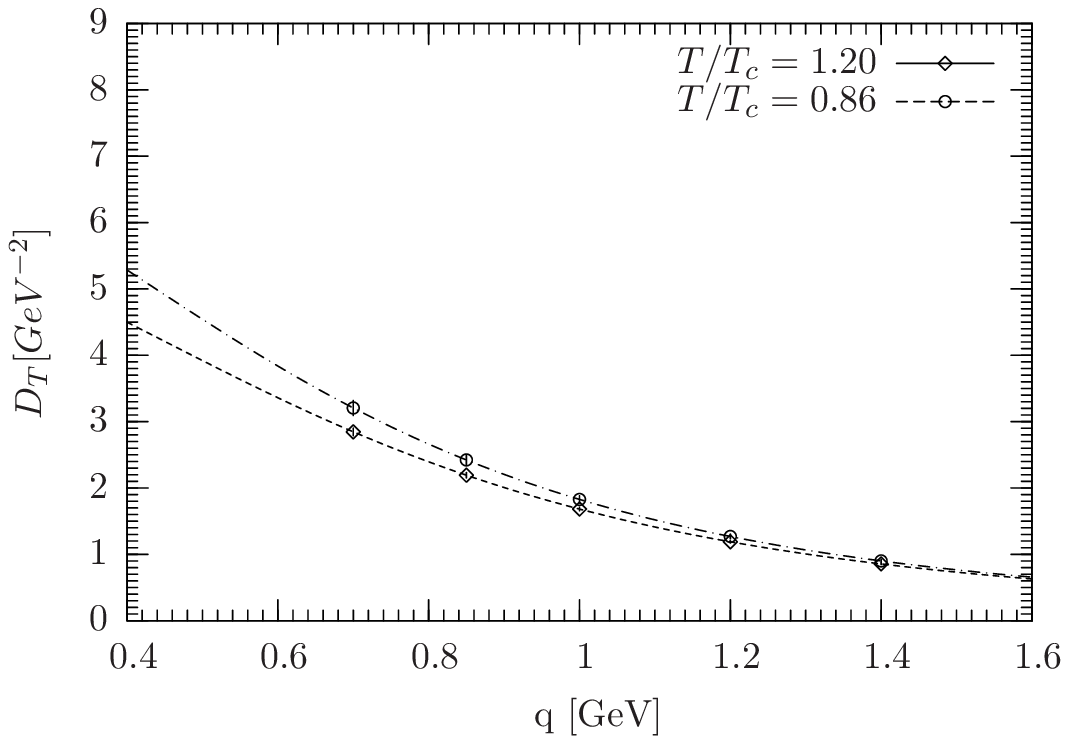} 
 }
\caption{Continuum extrapolated values of $D_L(q)$ (\LHS) and 
$D_T(q)$ (\RHS) together with their respective interpolation curves 
for two temperature values.}
\label{fig:fit_to_conti}
\end{figure*}

Our lattice propagator data obtained for $\beta=6.337$ as discussed in 
\Sec{sec:results} can now be compared with the values extrapolated 
to the continuum limit. This is shown in \Fig{fig:comp_to_conti}. 
In more detail, we can compare the continuum extrapolated values at 
some lower momentum -- say at $q=0.70 \,\mathrm{GeV}$ -- with those obtained 
from $a(\beta=6.337)=0.055 \,\mathrm{fm}$ and interpolated to 
the same momentum. Then we find deviations being smaller than 4\,\%. 
Thus, we are really justified to say that the results obtained for 
$\beta=6.337$ in the given momentum range are already very close to the 
continuum limit.
%------------------------------------------------------------------------------
%------------------------------------------------------------------------------
The continuum limit extrapolated propagators can be easily interpolated with 
formula \Eq{eq:stinglfit}. The results are drawn in \Fig{fig:fit_to_conti}.  

We conclude that for the higher $\beta$-values and the momentum range considered 
in this paper we are close to the continuum limit. Moreover, systematic effects 
as there are finite-volume and Gribov copy effects seem to be negligible for 
momenta above $0.8~\mathrm{GeV}$.  

%-------------------------------------------------------------------------------
%------------------------------------------------------------------------------

\section{Conclusions}
\label{sec:conclusions}
%----------------------
We have presented lattice results for the Landau gauge gluon and ghost
propagators computed in pure gauge $SU(3)$ lattice theory at non-zero
temperatures. 

Overall, our results agree with those published in
\cite{Fischer:2010fx} and there are hardly any deviations
which are not due to the lattice discretization or the finite
volume. However, our aim here was to go a step further and to
provide results in the continuum limit for temperatures below
and above the deconfinement phase transition, and this with
negligible systematic finite-volume and Gribov copy effects.
For this to become feasible we had to restrict the analysis to a
well-defined momentum range around $1 \,\mathrm{GeV}$.

The systematic effects were studied at two temperatures, $T = 0.86 T_c$ and
$T=1.20 T_c$, such that we cannot really tell, what happens very close to $T_c$.
Since for our reference value $\beta=6.337$ the critical temperature $T_c$
is reached with $N_{\tau}=12$, there is hope that also in this case we keep
close to the continuum limit and the other systematic effects are under
control. 

Our results and their parametrization can be further used to compare with 
the outcome of Dyson-Schwinger or functional renormalization group equations 
for the gluon propagators or employed as reliable input in Dyson-Schwinger 
studies of the quark propagator.
It is well-known that the non-perturbative continuum approaches rely 
on a truncated tower of equations for the propagators and vertex functions. 
The way of how it is truncated has a strong influence on the behavior,
especially at intermediate momenta.  Lattice results from first principles 
as those presented here can help to tune the truncation correspondingly.

Concentrating on this aim we have been forced to choose the lattice spacing $a$
and the linear spatial lattice extent $N_{\sigma}$ such that we were prevented
from going far towards the infrared limit. Therefore, we were not able to 
clarify the question, what the correct behavior is in the far infrared region.
Concerning this region we know, that the Gribov problem is serious and still 
not completely understood. This is also the reason, why we did not try in
this paper to give estimates for the color-electric and -magnetic screening 
masses.

Our results confirm that, contrary to the transverse gluon propagator $D_T$
and to the ghost propagator $G$, the longitudinal gluon propagator $D_L$ 
is sensitive to the deconfinement transition. 
However, despite of the fact that we are faced with a first order phase 
transition the response to it occurs relatively smooth. 
We were able to propose two parameters built from $D_L$ which 
can be employed as indicators (kind of ``order parameters'') for the 
transition itself. One might hope that they will help to shed additional 
light on the transition region also in full QCD. There, at least for 
$N_f=2$ quark flavours within the range of intermediate pion masses a 
rather smooth crossover is expected
(see e.g. \cite{DeTar:2011nm} and references therein, as well as 
\cite{Bornyakov:2011yb,Burger:2011zc}).  
We shall come back to this question in the near future.

\begin{samepage}
\section*{Acknowledgments}
%-------------------------
We acknowledge useful discussions with F.~Burger, C.~Fischer, A.~Maas, 
T.~Mendes, J.~Pawlowski, and L.~von~Smekal. R.~A. gratefully acknowledges 
financial support by the Yousef Jameel Foundation, A.~S.\ from the
European Reintegration Grant (FP7-PEOPLE-2009-RG No.256594),
V.K.M. and M.M.-P. by the Heisenberg-Landau program agreed upon between 
JINR and the German BMBF, V.B. by the Federal Programme Cadres of the 
Russian Ministry of Science and Education as well as by the grants 
RFBR 09-02-00338-a and NSh-6260.2010.2.
We thank the HLRN Berlin-Hannover and the JINR Computing Center
for generous supply with computing time. 
\end{samepage}
%------------------------------------------------------------------------------
\bibliographystyle{apsrev}
%\bibliography{citations_gluonghost}

\begin{thebibliography}{59}
\expandafter\ifx\csname natexlab\endcsname\relax\def\natexlab#1{#1}\fi
\expandafter\ifx\csname bibnamefont\endcsname\relax
  \def\bibnamefont#1{#1}\fi
\expandafter\ifx\csname bibfnamefont\endcsname\relax
  \def\bibfnamefont#1{#1}\fi
\expandafter\ifx\csname citenamefont\endcsname\relax
  \def\citenamefont#1{#1}\fi
\expandafter\ifx\csname url\endcsname\relax
  \def\url#1{\texttt{#1}}\fi
\expandafter\ifx\csname urlprefix\endcsname\relax\def\urlprefix{URL }\fi
\providecommand{\bibinfo}[2]{#2}
\providecommand{\eprint}[2][]{\url{#2}}

\bibitem[{\citenamefont{Hagedorn}(1970)}]{Hagedorn:1970gh}
\bibinfo{author}{\bibfnamefont{R.}~\bibnamefont{Hagedorn}},
  \bibinfo{journal}{Nucl.Phys.} \textbf{\bibinfo{volume}{B24}},
  \bibinfo{pages}{93} (\bibinfo{year}{1970}).

\bibitem[{\citenamefont{DeTar}(2011)}]{DeTar:2011nm}
\bibinfo{author}{\bibfnamefont{C.}~\bibnamefont{DeTar}} (\bibinfo{year}{2011}),
  \eprint{1101.0208}.

\bibitem[{\citenamefont{von Smekal et~al.}(1998)\citenamefont{von Smekal,
  Hauck, and Alkofer}}]{vonSmekal:1997vx}
\bibinfo{author}{\bibfnamefont{L.}~\bibnamefont{von Smekal}},
  \bibinfo{author}{\bibfnamefont{A.}~\bibnamefont{Hauck}}, \bibnamefont{and}
  \bibinfo{author}{\bibfnamefont{R.}~\bibnamefont{Alkofer}},
  \bibinfo{journal}{Ann. Phys.} \textbf{\bibinfo{volume}{267}},
  \bibinfo{pages}{1} (\bibinfo{year}{1998}), \eprint{hep-ph/9707327}.

\bibitem[{\citenamefont{Hauck et~al.}(1998)\citenamefont{Hauck, von Smekal, and
  Alkofer}}]{Hauck:1998fz}
\bibinfo{author}{\bibfnamefont{A.}~\bibnamefont{Hauck}},
  \bibinfo{author}{\bibfnamefont{L.}~\bibnamefont{von Smekal}},
  \bibnamefont{and} \bibinfo{author}{\bibfnamefont{R.}~\bibnamefont{Alkofer}},
  \bibinfo{journal}{Comput.Phys.Commun.} \textbf{\bibinfo{volume}{112}},
  \bibinfo{pages}{166} (\bibinfo{year}{1998}), \eprint{hep-ph/9804376}.

\bibitem[{\citenamefont{Roberts and Schmidt}(2000)}]{Roberts:2000aa}
\bibinfo{author}{\bibfnamefont{C.~D.} \bibnamefont{Roberts}} \bibnamefont{and}
  \bibinfo{author}{\bibfnamefont{S.~M.} \bibnamefont{Schmidt}},
  \bibinfo{journal}{Prog.Part.Nucl.Phys.} \textbf{\bibinfo{volume}{45}},
  \bibinfo{pages}{S1} (\bibinfo{year}{2000}), \eprint{nucl-th/0005064}.

\bibitem[{\citenamefont{Maris and Roberts}(2003)}]{Maris:2003vk}
\bibinfo{author}{\bibfnamefont{P.}~\bibnamefont{Maris}} \bibnamefont{and}
  \bibinfo{author}{\bibfnamefont{C.~D.} \bibnamefont{Roberts}},
  \bibinfo{journal}{Int.J.Mod.Phys.} \textbf{\bibinfo{volume}{E12}},
  \bibinfo{pages}{297} (\bibinfo{year}{2003}), \eprint{nucl-th/0301049}.

\bibitem[{\citenamefont{Gies}(2002)}]{Gies:2002af}
\bibinfo{author}{\bibfnamefont{H.}~\bibnamefont{Gies}},
  \bibinfo{journal}{Phys.Rev.} \textbf{\bibinfo{volume}{D66}},
  \bibinfo{pages}{025006} (\bibinfo{year}{2002}), \eprint{hep-th/0202207}.

\bibitem[{\citenamefont{Pawlowski et~al.}(2004)\citenamefont{Pawlowski, Litim,
  Nedelko, and von Smekal}}]{Pawlowski:2003hq}
\bibinfo{author}{\bibfnamefont{J.~M.} \bibnamefont{Pawlowski}},
  \bibinfo{author}{\bibfnamefont{D.~F.} \bibnamefont{Litim}},
  \bibinfo{author}{\bibfnamefont{S.}~\bibnamefont{Nedelko}}, \bibnamefont{and}
  \bibinfo{author}{\bibfnamefont{L.}~\bibnamefont{von Smekal}},
  \bibinfo{journal}{Phys. Rev. Lett.} \textbf{\bibinfo{volume}{93}},
  \bibinfo{pages}{152002} (\bibinfo{year}{2004}), \eprint{hep-th/0312324}.

\bibitem[{\citenamefont{Gribov}(1978)}]{Gribov:1977wm}
\bibinfo{author}{\bibfnamefont{V.~N.} \bibnamefont{Gribov}},
  \bibinfo{journal}{Nucl. Phys.} \textbf{\bibinfo{volume}{B139}},
  \bibinfo{pages}{1} (\bibinfo{year}{1978}).

\bibitem[{\citenamefont{Zwanziger}(2002)}]{Zwanziger:2001kw}
\bibinfo{author}{\bibfnamefont{D.}~\bibnamefont{Zwanziger}},
  \bibinfo{journal}{Phys. Rev.} \textbf{\bibinfo{volume}{D65}},
  \bibinfo{pages}{094039} (\bibinfo{year}{2002}), \eprint{hep-th/0109224}.

\bibitem[{\citenamefont{Zwanziger}(2004)}]{Zwanziger:2003cf}
\bibinfo{author}{\bibfnamefont{D.}~\bibnamefont{Zwanziger}},
  \bibinfo{journal}{Phys. Rev.} \textbf{\bibinfo{volume}{D69}},
  \bibinfo{pages}{016002} (\bibinfo{year}{2004}), \eprint{hep-ph/0303028}.

\bibitem[{\citenamefont{Kugo and Ojima}(1979)}]{Kugo:1979gm}
\bibinfo{author}{\bibfnamefont{T.}~\bibnamefont{Kugo}} \bibnamefont{and}
  \bibinfo{author}{\bibfnamefont{I.}~\bibnamefont{Ojima}},
  \bibinfo{journal}{Prog. Theor. Phys. Suppl.} \textbf{\bibinfo{volume}{66}},
  \bibinfo{pages}{1} (\bibinfo{year}{1979}).

\bibitem[{\citenamefont{Kugo}(1995)}]{Kugo:1995km}
\bibinfo{author}{\bibfnamefont{T.}~\bibnamefont{Kugo}} (\bibinfo{year}{1995}),
  \eprint{hep-th/9511033}.

\bibitem[{\citenamefont{Fischer et~al.}(2009)\citenamefont{Fischer, Maas, and
  Pawlowski}}]{Fischer:2008uz}
\bibinfo{author}{\bibfnamefont{C.~S.} \bibnamefont{Fischer}},
  \bibinfo{author}{\bibfnamefont{A.}~\bibnamefont{Maas}}, \bibnamefont{and}
  \bibinfo{author}{\bibfnamefont{J.~M.} \bibnamefont{Pawlowski}},
  \bibinfo{journal}{Annals Phys.} \textbf{\bibinfo{volume}{324}},
  \bibinfo{pages}{2408} (\bibinfo{year}{2009}), \eprint{0810.1987}.

\bibitem[{\citenamefont{Bornyakov et~al.}(2010)\citenamefont{Bornyakov,
  Mitrjushkin, and M{\"u}ller-Preussker}}]{Bornyakov:2009ug}
\bibinfo{author}{\bibfnamefont{V.~G.} \bibnamefont{Bornyakov}},
  \bibinfo{author}{\bibfnamefont{V.~K.} \bibnamefont{Mitrjushkin}},
  \bibnamefont{and}
  \bibinfo{author}{\bibfnamefont{M.}~\bibnamefont{M{\"u}ller-Preussker}},
  \bibinfo{journal}{Phys. Rev.} \textbf{\bibinfo{volume}{D81}},
  \bibinfo{pages}{054503} (\bibinfo{year}{2010}), \eprint{0912.4475}.

\bibitem[{\citenamefont{Bogolubsky et~al.}(2009)\citenamefont{Bogolubsky,
  Ilgenfritz, M{\"u}ller-Preussker, and Sternbeck}}]{Bogolubsky:2009dc}
\bibinfo{author}{\bibfnamefont{I.~L.} \bibnamefont{Bogolubsky}},
  \bibinfo{author}{\bibfnamefont{E.-M.} \bibnamefont{Ilgenfritz}},
  \bibinfo{author}{\bibfnamefont{M.}~\bibnamefont{M{\"u}ller-Preussker}},
  \bibnamefont{and}
  \bibinfo{author}{\bibfnamefont{A.}~\bibnamefont{Sternbeck}},
  \bibinfo{journal}{Phys. Lett.} \textbf{\bibinfo{volume}{B676}},
  \bibinfo{pages}{69} (\bibinfo{year}{2009}), \eprint{0901.0736}.

\bibitem[{\citenamefont{Heller et~al.}(1995)\citenamefont{Heller, Karsch, and
  Rank}}]{Heller:1995qc}
\bibinfo{author}{\bibfnamefont{U.~M.} \bibnamefont{Heller}},
  \bibinfo{author}{\bibfnamefont{F.}~\bibnamefont{Karsch}}, \bibnamefont{and}
  \bibinfo{author}{\bibfnamefont{J.}~\bibnamefont{Rank}},
  \bibinfo{journal}{Phys. Lett.} \textbf{\bibinfo{volume}{B355}},
  \bibinfo{pages}{511} (\bibinfo{year}{1995}), \eprint{hep-lat/9505016}.

\bibitem[{\citenamefont{Heller et~al.}(1998)\citenamefont{Heller, Karsch, and
  Rank}}]{Heller:1997nqa}
\bibinfo{author}{\bibfnamefont{U.~M.} \bibnamefont{Heller}},
  \bibinfo{author}{\bibfnamefont{F.}~\bibnamefont{Karsch}}, \bibnamefont{and}
  \bibinfo{author}{\bibfnamefont{J.}~\bibnamefont{Rank}},
  \bibinfo{journal}{Phys. Rev.} \textbf{\bibinfo{volume}{D57}},
  \bibinfo{pages}{1438} (\bibinfo{year}{1998}), \eprint{hep-lat/9710033}.

\bibitem[{\citenamefont{Cucchieri
  et~al.}(2001{\natexlab{a}})\citenamefont{Cucchieri, Karsch, and
  Petreczky}}]{Cucchieri:2000cy}
\bibinfo{author}{\bibfnamefont{A.}~\bibnamefont{Cucchieri}},
  \bibinfo{author}{\bibfnamefont{F.}~\bibnamefont{Karsch}}, \bibnamefont{and}
  \bibinfo{author}{\bibfnamefont{P.}~\bibnamefont{Petreczky}},
  \bibinfo{journal}{Phys. Lett.} \textbf{\bibinfo{volume}{B497}},
  \bibinfo{pages}{80} (\bibinfo{year}{2001}{\natexlab{a}}),
  \eprint{hep-lat/0004027}.

\bibitem[{\citenamefont{Cucchieri
  et~al.}(2001{\natexlab{b}})\citenamefont{Cucchieri, Karsch, and
  Petreczky}}]{Cucchieri:2001tw}
\bibinfo{author}{\bibfnamefont{A.}~\bibnamefont{Cucchieri}},
  \bibinfo{author}{\bibfnamefont{F.}~\bibnamefont{Karsch}}, \bibnamefont{and}
  \bibinfo{author}{\bibfnamefont{P.}~\bibnamefont{Petreczky}},
  \bibinfo{journal}{Phys. Rev.} \textbf{\bibinfo{volume}{D64}},
  \bibinfo{pages}{036001} (\bibinfo{year}{2001}{\natexlab{b}}),
  \eprint{hep-lat/0103009}.

\bibitem[{\citenamefont{Cucchieri et~al.}(2007)\citenamefont{Cucchieri, Maas,
  and Mendes}}]{Cucchieri:2007ta}
\bibinfo{author}{\bibfnamefont{A.}~\bibnamefont{Cucchieri}},
  \bibinfo{author}{\bibfnamefont{A.}~\bibnamefont{Maas}}, \bibnamefont{and}
  \bibinfo{author}{\bibfnamefont{T.}~\bibnamefont{Mendes}},
  \bibinfo{journal}{Phys. Rev.} \textbf{\bibinfo{volume}{D75}},
  \bibinfo{pages}{076003} (\bibinfo{year}{2007}), \eprint{hep-lat/0702022}.

\bibitem[{\citenamefont{Maas}(2010)}]{Maas:2009fg}
\bibinfo{author}{\bibfnamefont{A.}~\bibnamefont{Maas}}, \bibinfo{journal}{Chin.
  J. Phys.} \textbf{\bibinfo{volume}{34}}, \bibinfo{pages}{1328}
  (\bibinfo{year}{2010}), \eprint{0911.0348}.

\bibitem[{\citenamefont{Fischer et~al.}(2010)\citenamefont{Fischer, Maas, and
  M{\"u}ller}}]{Fischer:2010fx}
\bibinfo{author}{\bibfnamefont{C.~S.} \bibnamefont{Fischer}},
  \bibinfo{author}{\bibfnamefont{A.}~\bibnamefont{Maas}}, \bibnamefont{and}
  \bibinfo{author}{\bibfnamefont{J.~A.} \bibnamefont{M{\"u}ller}},
  \bibinfo{journal}{Eur. Phys. J.} \textbf{\bibinfo{volume}{C68}},
  \bibinfo{pages}{165} (\bibinfo{year}{2010}), \eprint{1003.1960}.

\bibitem[{\citenamefont{Cucchieri and Mendes}(2011)}]{Cucchieri:2011di}
\bibinfo{author}{\bibfnamefont{A.}~\bibnamefont{Cucchieri}} \bibnamefont{and}
  \bibinfo{author}{\bibfnamefont{T.}~\bibnamefont{Mendes}}
  (\bibinfo{year}{2011}), \eprint{1105.0176}.

\bibitem[{\citenamefont{Gruter et~al.}(2005)\citenamefont{Gruter, Alkofer,
  Maas, and Wambach}}]{Gruter:2004bb}
\bibinfo{author}{\bibfnamefont{B.}~\bibnamefont{Gruter}},
  \bibinfo{author}{\bibfnamefont{R.}~\bibnamefont{Alkofer}},
  \bibinfo{author}{\bibfnamefont{A.}~\bibnamefont{Maas}}, \bibnamefont{and}
  \bibinfo{author}{\bibfnamefont{J.}~\bibnamefont{Wambach}},
  \bibinfo{journal}{Eur.Phys.J.} \textbf{\bibinfo{volume}{C42}},
  \bibinfo{pages}{109} (\bibinfo{year}{2005}), \eprint{hep-ph/0408282}.

\bibitem[{\citenamefont{Maas et~al.}(2005)\citenamefont{Maas, Wambach, and
  Alkofer}}]{Maas:2005hs}
\bibinfo{author}{\bibfnamefont{A.}~\bibnamefont{Maas}},
  \bibinfo{author}{\bibfnamefont{J.}~\bibnamefont{Wambach}}, \bibnamefont{and}
  \bibinfo{author}{\bibfnamefont{R.}~\bibnamefont{Alkofer}},
  \bibinfo{journal}{Eur.Phys.J.} \textbf{\bibinfo{volume}{C42}},
  \bibinfo{pages}{93} (\bibinfo{year}{2005}), \eprint{hep-ph/0504019}.

\bibitem[{\citenamefont{Braun et~al.}(2010)\citenamefont{Braun, Gies, and
  Pawlowski}}]{Braun:2007bx}
\bibinfo{author}{\bibfnamefont{J.}~\bibnamefont{Braun}},
  \bibinfo{author}{\bibfnamefont{H.}~\bibnamefont{Gies}}, \bibnamefont{and}
  \bibinfo{author}{\bibfnamefont{J.~M.} \bibnamefont{Pawlowski}},
  \bibinfo{journal}{Phys. Lett.} \textbf{\bibinfo{volume}{B684}},
  \bibinfo{pages}{262} (\bibinfo{year}{2010}), \eprint{0708.2413}.

\bibitem[{\citenamefont{Fischer}(2009)}]{Fischer:2009wc}
\bibinfo{author}{\bibfnamefont{C.~S.} \bibnamefont{Fischer}},
  \bibinfo{journal}{Phys.Rev.Lett.} \textbf{\bibinfo{volume}{103}},
  \bibinfo{pages}{052003} (\bibinfo{year}{2009}), \eprint{0904.2700}.

\bibitem[{\citenamefont{Bornyakov and Mitrjushkin}(2010)}]{Bornyakov:2010nc}
\bibinfo{author}{\bibfnamefont{V.}~\bibnamefont{Bornyakov}} \bibnamefont{and}
  \bibinfo{author}{\bibfnamefont{V.}~\bibnamefont{Mitrjushkin}}
  (\bibinfo{year}{2010}), \eprint{1011.4790}.

\bibitem[{\citenamefont{Bornyakov and Mitrjushkin}(2011)}]{Bornyakov:2011jm}
\bibinfo{author}{\bibfnamefont{V.~G.} \bibnamefont{Bornyakov}}
  \bibnamefont{and} \bibinfo{author}{\bibfnamefont{V.~K.}
  \bibnamefont{Mitrjushkin}} (\bibinfo{year}{2011}), \eprint{1103.0442}.

\bibitem[{\citenamefont{Fabricius and Haan}(1984)}]{Fabricius:1984wp}
\bibinfo{author}{\bibfnamefont{K.}~\bibnamefont{Fabricius}} \bibnamefont{and}
  \bibinfo{author}{\bibfnamefont{O.}~\bibnamefont{Haan}},
  \bibinfo{journal}{Phys.Lett.} \textbf{\bibinfo{volume}{B143}},
  \bibinfo{pages}{459} (\bibinfo{year}{1984}).

\bibitem[{\citenamefont{Kennedy and Pendleton}(1985)}]{Kennedy:1985nu}
\bibinfo{author}{\bibfnamefont{A.}~\bibnamefont{Kennedy}} \bibnamefont{and}
  \bibinfo{author}{\bibfnamefont{B.}~\bibnamefont{Pendleton}},
  \bibinfo{journal}{Phys.Lett.} \textbf{\bibinfo{volume}{B156}},
  \bibinfo{pages}{393} (\bibinfo{year}{1985}).

\bibitem[{\citenamefont{Cabibbo and Marinari}(1982)}]{Cabibbo:1982zn}
\bibinfo{author}{\bibfnamefont{N.}~\bibnamefont{Cabibbo}} \bibnamefont{and}
  \bibinfo{author}{\bibfnamefont{E.}~\bibnamefont{Marinari}},
  \bibinfo{journal}{Phys. Lett.} \textbf{\bibinfo{volume}{B119}},
  \bibinfo{pages}{387} (\bibinfo{year}{1982}).

\bibitem[{\citenamefont{Necco and Sommer}(2002)}]{Necco:2001xg}
\bibinfo{author}{\bibfnamefont{S.}~\bibnamefont{Necco}} \bibnamefont{and}
  \bibinfo{author}{\bibfnamefont{R.}~\bibnamefont{Sommer}},
  \bibinfo{journal}{Nucl. Phys.} \textbf{\bibinfo{volume}{B622}},
  \bibinfo{pages}{328} (\bibinfo{year}{2002}), \eprint{hep-lat/0108008}.

\bibitem[{\citenamefont{Boyd et~al.}(1996)\citenamefont{Boyd, Engels, Karsch,
  Laermann, Legeland et~al.}}]{Boyd:1996bx}
\bibinfo{author}{\bibfnamefont{G.}~\bibnamefont{Boyd}},
  \bibinfo{author}{\bibfnamefont{J.}~\bibnamefont{Engels}},
  \bibinfo{author}{\bibfnamefont{F.}~\bibnamefont{Karsch}},
  \bibinfo{author}{\bibfnamefont{E.}~\bibnamefont{Laermann}},
  \bibinfo{author}{\bibfnamefont{C.}~\bibnamefont{Legeland}},
  \bibnamefont{et~al.}, \bibinfo{journal}{Nucl.Phys.}
  \textbf{\bibinfo{volume}{B469}}, \bibinfo{pages}{419} (\bibinfo{year}{1996}),
  \eprint{hep-lat/9602007}.

\bibitem[{\citenamefont{Parrinello and Jona-Lasinio}(1990)}]{Parrinello:1990pm}
\bibinfo{author}{\bibfnamefont{C.}~\bibnamefont{Parrinello}} \bibnamefont{and}
  \bibinfo{author}{\bibfnamefont{G.}~\bibnamefont{Jona-Lasinio}},
  \bibinfo{journal}{Phys. Lett.} \textbf{\bibinfo{volume}{B251}},
  \bibinfo{pages}{175} (\bibinfo{year}{1990}).

\bibitem[{\citenamefont{Zwanziger}(1990)}]{Zwanziger:1990tn}
\bibinfo{author}{\bibfnamefont{D.}~\bibnamefont{Zwanziger}},
  \bibinfo{journal}{Nucl. Phys.} \textbf{\bibinfo{volume}{B345}},
  \bibinfo{pages}{461} (\bibinfo{year}{1990}).

\bibitem[{\citenamefont{Bakeev et~al.}(2004)\citenamefont{Bakeev, Ilgenfritz,
  Mitrjushkin, and M{\"u}ller-Preussker}}]{Bakeev:2003rr}
\bibinfo{author}{\bibfnamefont{T.~D.} \bibnamefont{Bakeev}},
  \bibinfo{author}{\bibfnamefont{E.-M.} \bibnamefont{Ilgenfritz}},
  \bibinfo{author}{\bibfnamefont{V.~K.} \bibnamefont{Mitrjushkin}},
  \bibnamefont{and}
  \bibinfo{author}{\bibfnamefont{M.}~\bibnamefont{M{\"u}ller-Preussker}},
  \bibinfo{journal}{Phys. Rev.} \textbf{\bibinfo{volume}{D69}},
  \bibinfo{pages}{074507} (\bibinfo{year}{2004}), \eprint{hep-lat/0311041}.

\bibitem[{\citenamefont{Sternbeck et~al.}(2005)\citenamefont{Sternbeck,
  Ilgenfritz, M{\"u}ller-Preussker, and Schiller}}]{Sternbeck:2005tk}
\bibinfo{author}{\bibfnamefont{A.}~\bibnamefont{Sternbeck}},
  \bibinfo{author}{\bibfnamefont{E.-M.} \bibnamefont{Ilgenfritz}},
  \bibinfo{author}{\bibfnamefont{M.}~\bibnamefont{M{\"u}ller-Preussker}},
  \bibnamefont{and} \bibinfo{author}{\bibfnamefont{A.}~\bibnamefont{Schiller}},
  \bibinfo{journal}{Phys. Rev.} \textbf{\bibinfo{volume}{D72}},
  \bibinfo{pages}{014507} (\bibinfo{year}{2005}), \eprint{hep-lat/0506007}.

\bibitem[{\citenamefont{Bogolubsky et~al.}(2006)\citenamefont{Bogolubsky,
  Burgio, Mitrjushkin, and M{\"u}ller-Preussker}}]{Bogolubsky:2005wf}
\bibinfo{author}{\bibfnamefont{I.~L.} \bibnamefont{Bogolubsky}},
  \bibinfo{author}{\bibfnamefont{G.}~\bibnamefont{Burgio}},
  \bibinfo{author}{\bibfnamefont{V.~K.} \bibnamefont{Mitrjushkin}},
  \bibnamefont{and}
  \bibinfo{author}{\bibfnamefont{M.}~\bibnamefont{M{\"u}ller-Preussker}},
  \bibinfo{journal}{Phys. Rev.} \textbf{\bibinfo{volume}{D74}},
  \bibinfo{pages}{034503} (\bibinfo{year}{2006}), \eprint{hep-lat/0511056}.

\bibitem[{\citenamefont{Bogolubsky et~al.}(2008)\citenamefont{Bogolubsky,
  Bornyakov, Burgio, Ilgenfritz, Mitrjushkin, and
  M{\"u}ller-Preussker}}]{Bogolubsky:2007bw}
\bibinfo{author}{\bibfnamefont{I.~L.} \bibnamefont{Bogolubsky}},
  \bibinfo{author}{\bibfnamefont{V.~G.} \bibnamefont{Bornyakov}},
  \bibinfo{author}{\bibfnamefont{G.}~\bibnamefont{Burgio}},
  \bibinfo{author}{\bibfnamefont{E.-M.} \bibnamefont{Ilgenfritz}},
  \bibinfo{author}{\bibfnamefont{V.~K.} \bibnamefont{Mitrjushkin}},
  \bibnamefont{and}
  \bibinfo{author}{\bibfnamefont{M.}~\bibnamefont{M{\"u}ller-Preussker}},
  \bibinfo{journal}{Phys. Rev.} \textbf{\bibinfo{volume}{D77}},
  \bibinfo{pages}{014504} (\bibinfo{year}{2008}), \eprint{0707.3611}.

\bibitem[{\citenamefont{Nakamura and Plewnia}(1991)}]{Nakamura:1991ww}
\bibinfo{author}{\bibfnamefont{A.}~\bibnamefont{Nakamura}} \bibnamefont{and}
  \bibinfo{author}{\bibfnamefont{M.}~\bibnamefont{Plewnia}},
  \bibinfo{journal}{Phys. Lett.} \textbf{\bibinfo{volume}{B255}},
  \bibinfo{pages}{274} (\bibinfo{year}{1991}).

\bibitem[{\citenamefont{Bornyakov et~al.}(1993)\citenamefont{Bornyakov,
  Mitrjushkin, M{\"u}ller-Preussker, and Pahl}}]{Bornyakov:1993yy}
\bibinfo{author}{\bibfnamefont{V.~G.} \bibnamefont{Bornyakov}},
  \bibinfo{author}{\bibfnamefont{V.~K.} \bibnamefont{Mitrjushkin}},
  \bibinfo{author}{\bibfnamefont{M.}~\bibnamefont{M{\"u}ller-Preussker}},
  \bibnamefont{and} \bibinfo{author}{\bibfnamefont{F.}~\bibnamefont{Pahl}},
  \bibinfo{journal}{Phys. Lett.} \textbf{\bibinfo{volume}{B317}},
  \bibinfo{pages}{596} (\bibinfo{year}{1993}), \eprint{hep-lat/9307010}.

\bibitem[{\citenamefont{Bogolubsky et~al.}(1999)\citenamefont{Bogolubsky,
  Mitrjushkin, M{\"u}ller-Preussker, and Peter}}]{Bogolubsky:1999cb}
\bibinfo{author}{\bibfnamefont{I.~L.} \bibnamefont{Bogolubsky}},
  \bibinfo{author}{\bibfnamefont{V.~K.} \bibnamefont{Mitrjushkin}},
  \bibinfo{author}{\bibfnamefont{M.}~\bibnamefont{M{\"u}ller-Preussker}},
  \bibnamefont{and} \bibinfo{author}{\bibfnamefont{P.}~\bibnamefont{Peter}},
  \bibinfo{journal}{Phys. Lett.} \textbf{\bibinfo{volume}{B458}},
  \bibinfo{pages}{102} (\bibinfo{year}{1999}), \eprint{hep-lat/9904001}.

\bibitem[{\citenamefont{Bogolubsky et~al.}(2000)\citenamefont{Bogolubsky,
  Mitrjushkin, M{\"u}ller-Preussker, Peter, and Zverev}}]{Bogolubsky:1999ud}
\bibinfo{author}{\bibfnamefont{I.~L.} \bibnamefont{Bogolubsky}},
  \bibinfo{author}{\bibfnamefont{V.~K.} \bibnamefont{Mitrjushkin}},
  \bibinfo{author}{\bibfnamefont{M.}~\bibnamefont{M{\"u}ller-Preussker}},
  \bibinfo{author}{\bibfnamefont{P.}~\bibnamefont{Peter}}, \bibnamefont{and}
  \bibinfo{author}{\bibfnamefont{N.~V.} \bibnamefont{Zverev}},
  \bibinfo{journal}{Phys. Lett.} \textbf{\bibinfo{volume}{B476}},
  \bibinfo{pages}{448} (\bibinfo{year}{2000}), \eprint{hep-lat/9912017}.

\bibitem[{\citenamefont{Bali et~al.}(1995)\citenamefont{Bali, Bornyakov,
  M{\"u}ller-Preussker, and Pahl}}]{Bali:1994jg}
\bibinfo{author}{\bibfnamefont{G.~S.} \bibnamefont{Bali}},
  \bibinfo{author}{\bibfnamefont{V.}~\bibnamefont{Bornyakov}},
  \bibinfo{author}{\bibfnamefont{M.}~\bibnamefont{M{\"u}ller-Preussker}},
  \bibnamefont{and} \bibinfo{author}{\bibfnamefont{F.}~\bibnamefont{Pahl}},
  \bibinfo{journal}{Nucl. Phys. Proc. Suppl.} \textbf{\bibinfo{volume}{42}},
  \bibinfo{pages}{852} (\bibinfo{year}{1995}), \eprint{hep-lat/9412027}.

\bibitem[{\citenamefont{Bali et~al.}(1996)\citenamefont{Bali, Bornyakov,
  M{\"u}ller-Preussker, and Schilling}}]{Bali:1996dm}
\bibinfo{author}{\bibfnamefont{G.~S.} \bibnamefont{Bali}},
  \bibinfo{author}{\bibfnamefont{V.}~\bibnamefont{Bornyakov}},
  \bibinfo{author}{\bibfnamefont{M.}~\bibnamefont{M{\"u}ller-Preussker}},
  \bibnamefont{and}
  \bibinfo{author}{\bibfnamefont{K.}~\bibnamefont{Schilling}},
  \bibinfo{journal}{Phys. Rev.} \textbf{\bibinfo{volume}{D54}},
  \bibinfo{pages}{2863} (\bibinfo{year}{1996}), \eprint{hep-lat/9603012}.

\bibitem[{\citenamefont{Bogolubsky et~al.}(2007)\citenamefont{Bogolubsky,
  Bornyakov, Burgio, Ilgenfritz, Mitrjushkin, M{\"u}ller-Preussker, and
  Schemel}}]{Bogolubsky:2007pq}
\bibinfo{author}{\bibfnamefont{I.~L.} \bibnamefont{Bogolubsky}},
  \bibinfo{author}{\bibfnamefont{V.~G.} \bibnamefont{Bornyakov}},
  \bibinfo{author}{\bibfnamefont{G.}~\bibnamefont{Burgio}},
  \bibinfo{author}{\bibfnamefont{E.-M.} \bibnamefont{Ilgenfritz}},
  \bibinfo{author}{\bibfnamefont{V.~K.} \bibnamefont{Mitrjushkin}},
  \bibinfo{author}{\bibfnamefont{M.}~\bibnamefont{M{\"u}ller-Preussker}},
  \bibnamefont{and} \bibinfo{author}{\bibfnamefont{P.}~\bibnamefont{Schemel}},
  \bibinfo{journal}{PoS} \textbf{\bibinfo{volume}{LAT2007}},
  \bibinfo{pages}{318} (\bibinfo{year}{2007}), \eprint{0710.3234}.

\bibitem[{\citenamefont{Damm et~al.}(1998)\citenamefont{Damm, Kerler, and
  Mitrjushkin}}]{Damm:1998pd}
\bibinfo{author}{\bibfnamefont{G.}~\bibnamefont{Damm}},
  \bibinfo{author}{\bibfnamefont{W.}~\bibnamefont{Kerler}}, \bibnamefont{and}
  \bibinfo{author}{\bibfnamefont{V.~K.} \bibnamefont{Mitrjushkin}},
  \bibinfo{journal}{Phys. Lett.} \textbf{\bibinfo{volume}{B433}},
  \bibinfo{pages}{88} (\bibinfo{year}{1998}), \eprint{hep-lat/9802028}.

\bibitem[{\citenamefont{Sternbeck et~al.}(2006)\citenamefont{Sternbeck,
  Ilgenfritz, M{\"u}ller-Preussker, Schiller, and
  Bogolubsky}}]{Sternbeck:2006cg}
\bibinfo{author}{\bibfnamefont{A.}~\bibnamefont{Sternbeck}},
  \bibinfo{author}{\bibfnamefont{E.-M.} \bibnamefont{Ilgenfritz}},
  \bibinfo{author}{\bibfnamefont{M.}~\bibnamefont{M{\"u}ller-Preussker}},
  \bibinfo{author}{\bibfnamefont{A.}~\bibnamefont{Schiller}}, \bibnamefont{and}
  \bibinfo{author}{\bibfnamefont{I.~L.} \bibnamefont{Bogolubsky}},
  \bibinfo{journal}{PoS} \textbf{\bibinfo{volume}{LAT2006}},
  \bibinfo{pages}{076} (\bibinfo{year}{2006}), \eprint{hep-lat/0610053}.

\bibitem[{\citenamefont{Sternbeck}(2006)}]{Sternbeck:2006rd}
\bibinfo{author}{\bibfnamefont{A.}~\bibnamefont{Sternbeck}},
  \bibinfo{type}{{PhD thesis}}, \bibinfo{school}{Humboldt-University Berlin}
  (\bibinfo{year}{2006}), \eprint{hep-lat/0609016}.

\bibitem[{\citenamefont{Leinweber et~al.}(1999)\citenamefont{Leinweber,
  Skullerud, Williams, and Parrinello}}]{Leinweber:1998uu}
\bibinfo{author}{\bibfnamefont{D.~B.} \bibnamefont{Leinweber}},
  \bibinfo{author}{\bibfnamefont{J.~I.} \bibnamefont{Skullerud}},
  \bibinfo{author}{\bibfnamefont{A.~G.} \bibnamefont{Williams}},
  \bibnamefont{and}
  \bibinfo{author}{\bibfnamefont{C.}~\bibnamefont{Parrinello}}
  (\bibinfo{collaboration}{UKQCD}), \bibinfo{journal}{Phys. Rev.}
  \textbf{\bibinfo{volume}{D60}}, \bibinfo{pages}{094507}
  (\bibinfo{year}{1999}), \eprint{hep-lat/9811027}.

\bibitem[{\citenamefont{Stingl}(1996)}]{Stingl:1994nk}
\bibinfo{author}{\bibfnamefont{M.}~\bibnamefont{Stingl}},
  \bibinfo{journal}{Z.Phys.} \textbf{\bibinfo{volume}{A353}},
  \bibinfo{pages}{423} (\bibinfo{year}{1996}), \eprint{hep-th/9502157}.

\bibitem[{\citenamefont{Cucchieri et~al.}(2003)\citenamefont{Cucchieri, Mendes,
  and Taurines}}]{Cucchieri:2003di}
\bibinfo{author}{\bibfnamefont{A.}~\bibnamefont{Cucchieri}},
  \bibinfo{author}{\bibfnamefont{T.}~\bibnamefont{Mendes}}, \bibnamefont{and}
  \bibinfo{author}{\bibfnamefont{A.~R.} \bibnamefont{Taurines}},
  \bibinfo{journal}{Phys.Rev.} \textbf{\bibinfo{volume}{D67}},
  \bibinfo{pages}{091502} (\bibinfo{year}{2003}), \eprint{hep-lat/0302022}.

\bibitem[{\citenamefont{Dudal et~al.}(2008)\citenamefont{Dudal, Gracey,
  Sorella, Vandersickel, and Verschelde}}]{Dudal:2008sp}
\bibinfo{author}{\bibfnamefont{D.}~\bibnamefont{Dudal}},
  \bibinfo{author}{\bibfnamefont{J.~A.} \bibnamefont{Gracey}},
  \bibinfo{author}{\bibfnamefont{S.~P.} \bibnamefont{Sorella}},
  \bibinfo{author}{\bibfnamefont{N.}~\bibnamefont{Vandersickel}},
  \bibnamefont{and}
  \bibinfo{author}{\bibfnamefont{H.}~\bibnamefont{Verschelde}},
  \bibinfo{journal}{Phys. Rev.} \textbf{\bibinfo{volume}{D78}},
  \bibinfo{pages}{065047} (\bibinfo{year}{2008}), \eprint{0806.4348}.

\bibitem[{\citenamefont{Dudal et~al.}(2011)\citenamefont{Dudal, Sorella, and
  Vandersickel}}]{Dudal:2011gd}
\bibinfo{author}{\bibfnamefont{D.}~\bibnamefont{Dudal}},
  \bibinfo{author}{\bibfnamefont{S.}~\bibnamefont{Sorella}}, \bibnamefont{and}
  \bibinfo{author}{\bibfnamefont{N.}~\bibnamefont{Vandersickel}},
  \bibinfo{journal}{Phys.Rev.} \textbf{\bibinfo{volume}{D84}},
  \bibinfo{pages}{065039} (\bibinfo{year}{2011}), \eprint{1105.3371}.

\bibitem[{\citenamefont{Cucchieri and Mendes}(2007)}]{Cucchieri:2007md}
\bibinfo{author}{\bibfnamefont{A.}~\bibnamefont{Cucchieri}} \bibnamefont{and}
  \bibinfo{author}{\bibfnamefont{T.}~\bibnamefont{Mendes}},
  \bibinfo{journal}{PoS} \textbf{\bibinfo{volume}{LAT2007}},
  \bibinfo{pages}{297} (\bibinfo{year}{2007}), \eprint{0710.0412}.

\bibitem[{\citenamefont{Bornyakov et~al.}(2011)\citenamefont{Bornyakov,
  Horsley, Nakamura, Polikarpov, Rakow et~al.}}]{Bornyakov:2011yb}
\bibinfo{author}{\bibfnamefont{V.}~\bibnamefont{Bornyakov}},
  \bibinfo{author}{\bibfnamefont{R.}~\bibnamefont{Horsley}},
  \bibinfo{author}{\bibfnamefont{Y.}~\bibnamefont{Nakamura}},
  \bibinfo{author}{\bibfnamefont{M.}~\bibnamefont{Polikarpov}},
  \bibinfo{author}{\bibfnamefont{P.}~\bibnamefont{Rakow}}, \bibnamefont{et~al.}
  (\bibinfo{year}{2011}), \eprint{1102.4461}.

\bibitem[{\citenamefont{Burger et~al.}(2011)\citenamefont{Burger, Ilgenfritz,
  Kirchner, Lombardo, M{\"u}ller-Preussker et~al.}}]{Burger:2011zc}
\bibinfo{author}{\bibfnamefont{F.}~\bibnamefont{Burger}},
  \bibinfo{author}{\bibfnamefont{E.-M.} \bibnamefont{Ilgenfritz}},
  \bibinfo{author}{\bibfnamefont{M.}~\bibnamefont{Kirchner}},
  \bibinfo{author}{\bibfnamefont{M.}~\bibnamefont{Lombardo}},
  \bibinfo{author}{\bibfnamefont{M.}~\bibnamefont{M{\"u}ller-Preussker}},
  \bibnamefont{et~al.} (\bibinfo{year}{2011}), \eprint{1102.4530}.

\end{thebibliography}

\end{document}